\documentclass{aa}  
\usepackage{graphicx}  
\usepackage{subcaption}
\usepackage{amsmath,amssymb} 
\usepackage{natbib}
\usepackage{soul}
\usepackage{txfonts}
\usepackage{hyperref}
\begin{document}

   \title{Tracing outflows in the AGN forbidden region with SINFONI}
   \author{D. Kakkad\inst{1}, V. Mainieri\inst{1}, P. Padovani\inst{1}, G. Cresci\inst{2}, B. Husemann\inst{1}, S. Carniani\inst{2,3,4}, M. Brusa\inst{5,6}, A. Lamastra\inst{7}, G. Lanzuisi\inst{5,6}, E. Piconcelli\inst{7}, M. Schramm\inst{8,9} 
            }

   \institute{European Southern Observatory, Karl-Schwarzschild-Str. 2, 85748, Garching bei M\"unchen, Germany\\
              \email{dkakkad@eso.org}   \and
   INAF - Observatorio Astrofisico di Arcetri, Largo E. Fermi 5, I-50125, Firenze, Italy \and
   Cavendish Laboratory, University of Cambridge, Madingley Road, Cambridge CB3 0HA, UK \and
   Dipartimento di Fisica e Astronomia, Universit\`a di Firenze, Via G. Sansone 1, I-50019, Sesto Fiorentino (Firenze), Italy \and
   Dipartimento di Fisica e Astronomia, Universit\`{a} di Bologna, viale Berti Pichat 6/2, I-40127 Bologna, Italy \and
   INAF- Osservatorio Astronomico di Bologna, via Ranzani 1, I-40127 Bologna, Italy \and
   INAF- Osservatorio Astronomico di Roma, via Frascati 33, 00040 Monteporzio Catone, Italy \and
   Frequency Measurement Group, National Institute of Advanced Industrial Science and Technology (AIST) Tsukuba-central 3-1, Umezono 1-1-1, Tsukuba, Ibaraki 305-8563, Japan \and
  National Astronomical Observatory of Japan, 2-21-1 Osawa, Mitaka, Tokyo 181-8588, Japan 
}
   \date{Received ?; accepted ?=}
   
% \abstract{}{}{}{}{} 
% 5 {} token are mandatory
 
  \abstract
  {AGN driven outflows are invoked in numerical simulations to
    reproduce several observed properties of local galaxies.  The z >
    1 epoch is of particular interest as it was during this time that
    the volume averaged star formation and the accretion rate of black
    holes were maximum. Radiatively driven outflows are therefore
    believed to be common during this epoch.}
  {We aim to trace and characterize outflows in AGN hosts with high
    mass accretion rates at z > 1 using integral field
    spectroscopy. We obtain spatially-resolved kinematics of the
    [\ion{O}{iii}] $\lambda$5007 line in two targets which reveal the morphology and
    spatial extension of the outflows.}
  {We present J and H+K band SINFONI observations of 5 AGNs at 1.2 < z
    < 2.2. To maximize the chance of observing radiatively driven
    outflows, our sample was pre-selected based on peculiar values of
    the Eddington ratio and the hydrogen column density of the
    surrounding interstellar medium. We observe high velocity
    ($\sim$600-1900 km/s) and kiloparsec scale extended ionized outflows in
    at least 3 of our targets, using [\ion{O}{iii}] $\lambda$5007 line
    kinematics tracing the AGN narrow line region. We estimate the
    total mass of the outflow, the mass outflow rate, and the kinetic
    power of the outflows based on theoretical models and report on
    the uncertainties associated with them.}
  {We find mass outflow rates of $\sim 1 - 10~M_{\odot}$/yr for
      the sample presented in this paper. Based on the high star
    formation rates of the host galaxies, the observed outflow kinetic
    power and the expected power due to the AGN, we infer that both
    star formation and AGN radiation could be the dominant source for
    the outflows. The outflow models suffer from large uncertainties,
    hence we call for further detailed observations for an accurate
    determination of the outflow properties to confirm the exact
    source of these outflows.}  {} 

\authorrunning{Kakkad et al.} 
   \maketitle
%
%________________________________________________________________

\section{Introduction} \label{sec_introduction}

It is a well established fact that most galaxies in the Universe host
a super massive black hole (SMBH) in their nucleus \citep{magorrian98,
  kormendy11}.  These black holes grow by accretion of surrounding gas
and dust \citep{silk98} and may turn active for a certain period of
time ($\sim$10$^{5}$-10$^{7}$ yrs) \citep{martini01,schawinski15,
  king15}. Various galaxy evolutionary models invoke outflows driven
by active galactic nuclei (AGN) to reproduce several properties of
local massive galaxies
\citep{silk98,granato04,matteo05,croton06,hopkins06,menci06,fabian12,king_pounds15}. Such
outflows couple to the surrounding gas and dust and this process,
named AGN feedback, is invoked to explain various observed properties
such as the black hole and bulge mass relation and the exponential
break in the galaxy luminosity function, to name a few \citep{silk12,
  kormendy13}.

The coupling between the outflows and the interstellar medium (ISM)
could be in the form of mechanical energy, commonly called jet-mode
feedback or radiation energy, called the radiative-mode feedback (see
\citealt{fabian12} and \citealt{heckman14} for a recent review).  Jet-mode
feedback occurs in black holes having low mass accretion rates. The
outflows from such black holes are in the form of relativistic jets
with narrow opening angles launched along the axis of the accretion
disc. The impact of such feedback mode has been confirmed through
X-ray observations of the centres of galaxy clusters or groups with a
radio-loud AGN in their centres \citep{cavagnolo11, david11,
  nesvadba08, nesvadba11}. On the other hand, black holes with high
mass accretion rates are more likely to drive the radiative feedback
mode since these radiative winds are believed to originate from the
accretion disc \citep{granato04, matteo05, menci08,
  nayakshin14}. Although there is some observational evidence for the
presence of radiative feedback in a few objects \citep{cano-diaz12,zakamska15,cresci15, perna15, carniani15}, we are far from reaching
a general conclusion on its impact on the host galaxy.

Outflows have been commonly revealed in local as well as high redshift
galaxies using X-ray and UV absorption line studies with velocities
reaching >1000 km/s
\citep{crenshaw99,chartas02,ganguly07,piconcelli05,tombesi10}.  Since
the radiative mode of feedback should be more relevant at 
$1< z <3$, the epoch of peak volume averaged accretion density of the
black holes and the star formation density of the galaxies
\citep{shankar09, madau14}, there has been an increasing interest to
observe galaxies in this redshift range to detect such outflows. In
order to quantify their impact on the host galaxy one needs to
determine their spatial extension and energetics. Long slit
spectroscopy has been widely used to detect such outflows from the
presence of broad and extended emission line profiles in the object
spectra \citep{alexander10, harrison12, brusa15a}. However, one
dimensional spectroscopy has the disadvantage that the spatial
information and the outflow morphology cannot be inferred from
it. Integral field spectroscopy (IFS) is the ideal tool as one can get
an idea of both the spatial extension and the morphology of the
outflow around the host galaxy in addition to obtaining the total gas
content traced by the respective emission lines. In recent years,
there has been extensive work on this front using IFS on local as well
as high redshift quasars. A few examples of such works are described
in brief below.

\cite{harrison14} used GMOS-IFU (Gemini Multi Object
Spectrograph-Integral Field Unit) observations to spatially resolve
ionized gas kinematics in a sample of 16 local radio-quiet luminous
Type 2 AGN. They found high velocity and disturbed gas extended over
scales of the host galaxies in all of their objects. Though no
specific mechanism behind such outflows, i.e. star formation driven or
AGN driven, is favored for the sample in general, the most extreme
ionized gas velocities seem to be due to the AGN. \cite{husemann13}
also studied the gas kinematics of a sample of $\sim$30 low redshift
QSOs using Potsdam Multi-Aperture Spectrophotometer (PMAS) where the
disturbed gas kinematics is attributed to small scale radio-jet and
cloud interactions rather than being AGN-driven.  \cite{cano-diaz12}
found an extended kiloparsec scale quasar driven outflow for a high
redshift Type 1 quasar using Spectrograph for INtegral Field in the
Near Infrared (SINFONI) data. The outflow was asymmetric in morphology
and the star formation, traced by the narrow component of H$\alpha$
line, was mostly found in the regions not directly affected by the
strong outflow. This was one of the first direct observational
evidence of a negative AGN-feedback. \cite{cresci15} also detected an
extended outflow for a high redshift Type 2 quasar. However, in this
case, the outflow seems to affect the distribution of star formation
in the host galaxy such that star formation is suppressed in the
regions dominated by ionized outflows, but enhanced at the edges of
the outflow, making it one of the first observational evidences of
both negative and positive feedback at play in a galaxy. Finally,
\cite{carniani15} studied a sample of six high redshift luminous
quasars and found extended kiloparsec scale and high velocity outflows
in all their objects. These studies demonstrate the capability of IFS
to investigate the impact of AGN on host galaxies, both at high as
well as low redshifts. They also suggest that outflows are very common
in this redshift range, which should be ideal to study AGN feedback
due to radiation pressure driven outflows.

In most of these works, the key diagnostic feature for the presence of
kiloparsec scale outflows is the presence of asymmetric
[\ion{O}{iii}] $\lambda$5007 profiles. [\ion{O}{iii}] $\lambda$4959,5007 are ideal
tracers of ionized gas in the narrow line region (NLR) as it could not
be emitted from the high density sub-parsec scales of the broad line
region (BLR). The direction of the outflow can be inferred from the
presence of a blue or a red wing in the asymmetric [\ion{O}{iii}] $\lambda$5007
profile, which indicates gas flow towards or away from the observer
respectively. Assuming a bi-conical outflow morphology in most
galaxies, it is not uncommon to observe only the blue wing as the red
wing is thought to be obscured by dust in the host galaxy.

IFS observations, as the ones listed above, are very expensive in
terms of telescope time, and therefore previous studies have always
tried to pre-select the targets in order to maximize the chances of
observing the AGN in an outflowing phase \citep{lipari06,
  brusa15a}. We have recently completed a SINFONI program at VLT on a
sample of five radio quiet Quasi-stellar objects (QSOs) at
$1.2 < z < 2.2$.  The main goal of this program was to prove the
effectiveness of selecting AGN in an outflowing phase based on the
peculiar values of the Eddington ratio (which is the ratio of the
bolometric luminosity and the Eddington luminosity) or its mass
accretion rate and column density of the surrounding interstellar
medium. We selected only radio-quiet QSOs since the main goal was to
use this selection criterion for studies on the impact of radiative
feedback on the host galaxy.

This paper is arranged as follows: in Sec. \ref{sec_sample_selection},
we present the selection strategy of our sample, in
Sec. \ref{sec_observations}, we discuss the observations and the
technical details of the data reduction procedure,
Sec. \ref{sec_data_analysis} presents a detailed description of line
fitting, creation of kinematic maps and the velocity definitions used
in the paper. In Sec. \ref{sec_results}, we describe the properties of
individual objects derived from the line fitting and the kinematic
maps. In Sec. \ref{sec_outflow_properties}, we provide details about
the formulas used and the assumptions that go into our model while
deriving the outflow properties. In Sec. \ref{sec_discussion} we
discuss our results and compare them with previous work and finally
the conclusions are presented in
Sec. \ref{sec_conclusions}. Throughout this paper, we use an $H_{o}$ =
70 km s$^{-1}$, $\Omega_{\Lambda}$ = 0.7 and $\Omega_{M}$ = 0.3
cosmology.

%__________________________________________________________________

\section{Sample selection}   \label{sec_sample_selection}

\begin{figure}
\centering
\includegraphics[width=9cm]{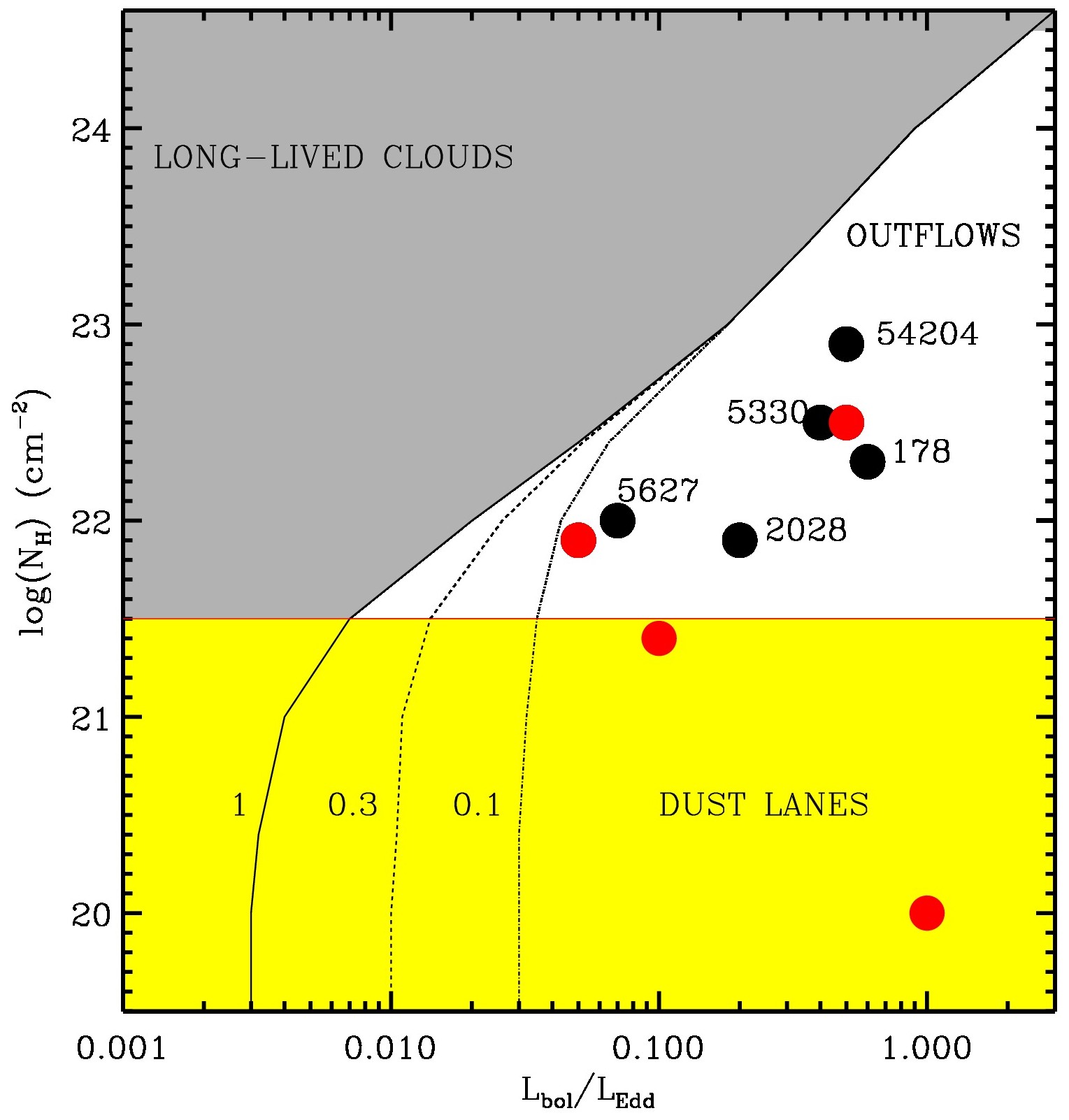} 
\caption{Our sample from the COSMOS field in the hydrogen column
  density, $\mathrm{N_H}$ vs. Eddington ratio, $\lambda$ =
  $\mathrm{L_{bol}/L_{Edd}}$ plot, adapted from \cite{fabian08}. The
  data points from this work are depicted in black circles, while
  those in red circles have been obtained from \cite{brusa15a} for the
  objects for which the presence of outflows have been verified (see
  Sec. \ref{sec_discussion}). Various regions of the plot are divided
  by the expected Eddington limit curves shown in solid, dashed and
  dotted lines representing galactic dust abundance, 0.3 and 0.1 of
  the galactic dust abundance respectively. The objects in grey
  regions are expected to have long lived clouds, while those on the
  un-shaded region are expected to be in a "blow-out" or an outflowing
  phase.  \label{forbidden_region}}
\end{figure}

As mentioned before, to maximize the chance of observing outflows
driven by an AGN, we need to pre-select our objects based on peculiar
values of the physical properties of the black hole such as its mass
accretion rate or the Eddington ratio and the column density of the
surrounding interstellar medium (ISM). Radiatively driven winds are
believed to originate from the acceleration of the disk outflows by
the AGN radiation field \citep{begelman03}. Therefore, our selection
criterion is skewed towards objects showing high mass accretion rates
or equivalently objects having higher Eddington ratio.  An object at
higher Eddington ratio, will have a tendency to induce a larger
radiation pressure on the surrounding ISM. The additional constraint
on the column density is motivated by the impact of the radiation
pressure generated by the SMBH on the cold gas responsible for the
nuclear obscuration \citep{fabian08}. The ISM might be able to
withstand the high radiation pressure from the AGN provided it has
enough gravitational support, an estimate of which can be obtained
from the hydrogen column density measurements. The coupling area
between the radiation pressure and that of the surrounding gas is
given by the cross section of the particles and in presence of a gas
(say a dominant atom like hydrogen), this would be simply the Thomson
cross section. However, the ISM consists of both gas and dust grains and in the presence of dust radiation pressure is more efficient due to its higher cross section, thereby lowering the Eddington limit. The effective Eddington limit
due to interaction with dust, defined at the balance between radiation
pressure and gravity, can be a factor of 1000 lower than the
``classical'' $\mathrm{L_{Edd}}$ for a gas with a Galactic dust-to-gas
ratio exposed to a typical quasar spectrum. The result is that long
lived clouds would avoid a region of intermediate column densities and
high Eddington ratios. Fig. \ref{forbidden_region} shows the possible
dividing lines between the objects having long-lived clouds (or in
other words, those which are not expected to show outflows) shown by
the grey region and those which are expected to be in an active
outflowing phase (or the "forbidden region" for long lived clouds), shown by the un-shaded region labeled as "Outflows", adapted from \cite{fabian08}. The curve labeled "1" in
Fig. \ref{forbidden_region} shows the effective Eddington limit for a
standard dust abundance for a galaxy while the dashed and dotted lines
are for dust abundance of 0.3 and 0.1 of galactic dust abundance
respectively.

We looked for objects in this forbidden region from the full
XMM-COSMOS catalogue \citep{scoville07,hasinger07, cappelluti09} which
consists of $\sim$1800 AGN from the entire 2 deg$^{2}$ COSMOS field,
with data having a wavelength coverage from radio to UV, with
additional information on the morphology, stellar masses, star
formation rate and infrared (IR) luminosities of the host galaxies
\citep{brusa10,salvato11,civano12,bongiorno12,rosario12}.  We restrict
our study to objects at z>1 because it is at these redshifts that we
expect the radiative feedback to have more significant impact. Our
parent sample consisted of 49 objects with z=1.2-2.2, having secure
spectroscopic redshifts, black hole mass and bolometric luminosity
measurements (to measure the Eddington ratio, $\lambda$ =
$\mathrm{L_{bol}/L_{Edd}}$) and reliable column density. The column
densities have been derived from detailed X-ray spectral fits
\citep{mainieri07,lanzuisi13}, the bolometric luminosities have been
estimated from SED fitting \citep{lusso11} and the black hole masses
from broad MgII lines in the optical spectra \citep{merloni10,
  matsuoka13} using the virial theorem. Since we want to detect the
presence of radiative outflows, we exclude 6 objects which are radio
loud with radio-to-optical flux density ratio (R =
log($\mathrm{S_{1.4 GHz}/S_{V}}$)) $> 1.4$. Of the remaining 43
objects which are radio-quiet, 20 are in the forbidden region. Of
these, we followed up the five best candidates in terms of the
location of the [\ion{O}{iii}] $\lambda$4959,5007 emission lines compared to
the OH sky emission lines in the infrared spectra. To estimate the
stellar mass of the host galaxies, we used an SED fitting technique to
model the observed photometry with a galactic and an AGN component
\citep{bonzini13}. We took advantage of the superb photometry up to
the Herschel bands to have a good estimate of the far-infrared (FIR)
emission which we used as a tracer of the star-formation
\citep{bonzini15}. The basic properties of the selected objects are
listed in Table \ref{sample_selection}. The results of the analysis of
XID2028, one of the objects from our selection sample, were published
in \cite{cresci15}. In this paper, we will present some of the
information from XID2028 data which were not reported in
\cite{cresci15} and the analysis of the rest of the sample from our
selection.

%______________________________________________________________
\begin{table*}
\centering          
\begin{tabular}{c c c c c c c c c c c}     % 9 columns 
\hline\hline       
Object ID & RA & DEC & z & log($\mathrm{L_{bol}}$) & log($\mathrm{m_{bh}}$) & log($\mathrm{N_{H}}$) & $\mathrm{M_{acc}}$ & $\mathrm{L/L_{edd}}$ & Log M* & SFR \\ 
 & (h:m:s) & $^{\circ}$:$^{\prime}$:$^{\prime\prime}$ & & (erg/s) & ($\mathrm{M_{\odot}}$) & (cm$^{-2}$) & (M$_{\odot}$/yr) & & (M$_{\odot}$) & (M$_{\odot}$/yr)\\
(1) & (2) & (3) & (4) & (5) & (6) & (7) & (8) & (9) & (10) & (11)\\
\hline                    
   178   & 10 00 14 & 02 28 37 & 1.253 & 45.7 & 7.78 & 22.3 & 0.8 & 0.6  & 10.8 & 134 \\ 
   5330  & 09 59 30 & 02 41 26 & 2.169 & 45.9 & 8.22 & 22.5 & 1.2 & 0.4  & 10.5 & 101\\
   5627  & 09 58 44 & 01 43 09 & 1.337 & 45.7 & 8.77 & 22.0 & 0.8 & 0.07 & 10.4 & 157\\
   54204 & 09 58 20 & 02 03 02 & 1.356 & 46.1 & 8.27 & 22.9 & 1.7 & 0.5  & 10.7 & 438\\
   2028  & 10 02 11 & 01 37 07 & 1.592 & 46.3 & 8.83 & 21.9 & 2.8 & 0.2  & 11.6 & 250\\
\hline                  
\end{tabular}
\caption{Our sample from the COSMOS field. Column (1): The X-ray ID of the objects; columns (2) and (3): Optical coordinates (J200); Column (4): the spectroscopic redshift of the targets; column (5) Bolometric luminosity from SED fitting; column (6) Black hole mass from broad MgII lines; column (7): column density from X-ray spectral fits; column (8): Mass accretion rate of the black hole; column (9): the Eddington ratio of the black hole; column (10): stellar mass of the host galaxy and column (11): star formation rate of the host galaxy following the procedure by \citealt{bonzini15}. See Sec. \ref{sec_sample_selection} for a detailed description of the selection procedure. \label{sample_selection}}
\end{table*}

\begin{table*}
\centering          
\begin{tabular}{c c c c c c}    
\hline\hline       
Object ID & $\mathrm{\lambda_{narrow}}$ & $\mathrm{v_{narrow}}$ & $\mathrm{\lambda_{broad}}$ & $\mathrm{v_{broad}}$ & $\mathrm{\Delta v}$\\ 
& ($\mu$m) & (km/s) & ($\mu$m) & (km/s) & (km/s)\\
(1) & (2) & (3) & (4) & (5) & (6)\\
\hline                    
   178   & 1.1310 $\pm$ 0.0001 & 308 $\pm$ 4  & 1.1299 $\pm$ 0.0001 & 1250 $\pm$ 130  & -275 $\pm$ 46\\  
   5330  & 1.5930 $\pm$ 0.0001 & 557 $\pm$ 21  & -                   & -              & -   \\   
   5627  & 1.1766 $\pm$ 0.0001 & 206 $\pm$ 66 & 1.1765 $\pm$ 0.0001 & 537  $\pm$ 50   & -9 $\pm$ 4\\
   54204 & 1.1810 $\pm$ 0.0002 & 305 $\pm$ 71 & 1.1796 $\pm$ 0.0004 & ~~1796 $\pm$ 882 & -359 $\pm$ 134\\
   2028* & 1.2989 $\pm$ 0.0001 & 366 $\pm$ 3  & 1.2976 $\pm$ 0.0001 & 611  $\pm$ 48   & -300 $\pm$ 50\\
\hline                  
\end{tabular}
\caption{[\ion{O}{iii}] $\lambda$5007 line fitting parameters for the J band integrated spectra of XID178, XID5627, XID54204 and XID2028 and H band spectra of XID5330. Details of the line fitting and the contraints are mentioned in Sec. \ref{sec_data_analysis}. (1) The X-ray ID of our sample; (2) the central wavelength of the narrow Gaussian component of [\ion{O}{iii}] $\lambda$5007 line; (3) the velocity corresponding to the width (FWHM) of the narrow Gaussian component; (4) the central wavelength of the broad Gaussian component of [\ion{O}{iii}] $\lambda$5007 line; (5) the velocity corresponding to the width (FWHM) of the broad Gaussian component; (6) velocity offset between the 
centroids of the narrow and the broad Gaussian components. \label{all_linefit_parameters}}
\tablefoot{*The values for XID2028 reported are from our line fitting results using the two Gaussian components for consistency.}
\end{table*}

\section{Observations and data reduction}   \label{sec_observations}

The aim of our observations is to confirm the presence of outflowing
winds in such short-lived objects, and spatially resolve outflows on
the scale of the host galaxy i.e. several kpcs. As mentioned before,
the key diagnostic feature used to detect outflowing gas is the
presence of broad and extended forbidden [\ion{O}{iii}]  lines at $\lambda$ =
5006.8 $\AA$. This traces the ionized gas in the narrow line region
(NLR), which could be extended up to the scale of the host galaxy.

At the redshift of our objects, the [\ion{O}{iii}] $\lambda$4959,5007 lines
fall in the near-infrared (NIR) and therefore we used SINFONI
\citep{eisenhauer03} at the UT4, Very Large Telescope
(VLT). Observations were taken between January-March, 2014 as part of
the program 092.A-0144 (PI Mainieri). The observations were carried
out in seeing limited mode in J band (targets with X-ray ID (XID):
XID178, XID5627, XID54204) providing a spectral resolution of 2000 and
H+K band (XID5330) providing a spectral resolution of 1500. SINFONI
provides rectangular pixels with a spatial scale of
0.125"$\times$0.25", which is re-sampled to a square pixel with a
spatial scale of 0.125"$\times$0.125" and a total field of view (FOV)
of 8"$\times$8". This means that the gas kinematics can be traced to a
spatial scale of few tens of kpc for the entire FOV for galaxies at
$z>1$. In the observations, the object was dithered by 3.5" across the
field of view to ensure an optimal sky subtraction during the data
reduction process without losing observing time. The total integration
time on source for XID178, XID5627 and XID54204 was about 3.5 hours
while for XID5330 it was about 50 minutes reaching depths of 9.67, 2.84, 4.96 and 1.29 $\mathrm{\times 10^{-15} erg~ s^{-1} cm^{-2} \mu m^{-1}}$ respectively. The air mass for all the
observation blocks was between X$\sim$ 1.1-1.2 for XID178, X$\sim$
1.1-1.4 for XID5627, X$\sim$ 1.2-1.4 for XID5330 and X$\sim$ 1.7-2.0
for XID54204. The standard stars for telluric and point spread
function (PSF) estimation were observed shortly before and after each
observing block with an air mass within 0.2 of that of the science
observations.

The data were reduced using the ESO-SINFONI pipeline (version 2.5.2)
which corrects for bad pixels and distortions, applies a flat field
and performs a wavelength calibration. The final image of the object
is reconstructed in the form of 32 slices which contains both the
spatial and the spectral information. The raw science frames were
first corrected for cosmic ray features using the Laplacian Cosmic Ray
Identification procedure (L. A. Cosmic) by \cite{vandokkum01} before
being fed into the pipeline. The sky subtraction was done externally
using the improved procedure proposed by \cite{davies07}. To remove
the infrared sky background, adjacent science frames were used as sky
for the object. Two consecutive frames are separated by $\sim$10
minutes, a time interval over which the infrared sky should not change
significantly. The final science frames were obtained after correcting
for telluric features and flux calibrating the cube using the standard
star observed before or after each observing blocks. The header
information was used to combine different science cubes within the
same observing block while the flux calibrated cubes of different
observing blocks were combined by measuring the offset in the centroid
emission in a given spectral channel.

%_______________________________________________________________

\section{Data analysis}   \label{sec_data_analysis}

For each of the final science cubes, the integrated spectrum was
extracted from a circular region, centred at the object position and
with a radius that maximizes the signal-to-noise ratio (S/N) of the
[\ion{O}{iii}] $\lambda$5007 line. Similar to previous studies
(e.g. \citealt{harrison14, perna15}) we performed a simultaneous fit
of the continuum, [\ion{O}{iii}] $\lambda$5007 and [\ion{O}{iii}] $\lambda$4959 using
the IDL routine MPFIT \citep{markwardt12}. The H$\beta$ line remained
undetected or it was very faint for all our objects.  The errors were
estimated from the standard deviation of the spectrum extracted from
an object-free region of the SINFONI field of
view. Eq. \ref{eq_fitting} below shows the mathematical form of the
function used for line fitting:

\begin{equation}
\label{eq_fitting}
f(x) = a\cdot e^{-b\cdot x} + \xi_{1}(x) + \xi_{2}(x)
\end{equation}

\noindent
where the first term corresponds to the local continuum with $a$ and
$b$ as free parameters. For fitting the overall [\ion{O}{iii}] $\lambda$5007
profile, we used either a single or a double Gaussian, the choice of
which depended on the S/N of the data and the ability of the
components to reproduce the emission line profile.  $\xi_{1}$ and
$\xi_{2}$ in Eq. \ref{eq_fitting} correspond to these two Gaussian
components - a narrow and a broad component, each with three free
parameters defining the central wavelength, width and the flux. The
narrow component corresponds to the systemic line which is at rest
with respect to the host galaxy, while the broad component(s) would
trace the outflowing or the turbulent gas. For all the line profiles,
the narrow and broad line widths were constrained. For the narrow
Gaussian component, we assume a maximum width (FWHM) of $\sim$500 km/s
corresponding to the rotational velocity of the host galaxy and motion
in the NLR while the width of the broad Gaussian was kept greater than
500 km/s to decouple it from the narrow component. When only a single
Gaussian was used no constrains on the line width were applied. The
line fitting was checked by plotting the residuals of the fit over the
[\ion{O}{iii}] $\lambda$5007 profile. The parameters of [\ion{O}{iii}] $\lambda$4959
were coupled to those of [\ion{O}{iii}] $\lambda$5007, imposing the same
velocity dispersion and a flux ratio of
$\mathrm{f_{[\ion{O}{iii}] 5007}/f_{[\ion{O}{iii}] 4959}} \approx$ 3 since they are
emitted from the same gas \citep{storey00, dimitrijevic07}.  The line
fit results of the integrated spectrum for each object are given in
Table \ref{all_linefit_parameters} where we list the wavelength
($\mu$m) and width (FWHM in km/s) of the individual Gaussian
components and the velocity offset between their centroids.

Since we are interested in spatially resolved kinematics, similar line
fitting procedure was performed across the entire field of view for
each spaxel separately. The robustness of the fitting procedure across
each pixel was checked by creating a residual map of the entire
[\ion{O}{iii}] $\lambda$5007 profile. From the results of the spaxel by spaxel
fit, we constructed flux maps. Wherever necessary, a re-binning was
performed on the cube to improve the S/N across each spaxel. Clearly,
this increases the spatial scale of each pixel in the field of view.

Wherever possible, for a better quantitative estimate of the spatial extension of the emission,  we created maps
  for the narrow and the broad components of the [\ion{O}{iii}] $\lambda$5007
  line profile separately, tracing the gas at rest with respect to the
  system and the outflowing gas. To verify if the emission is extended, we derived the surface
  brightness profiles of the narrow and the broad components as a
  function of radius from the peak of the flux weighted mean of the
  maps. These were compared to the surface brightness profile of the
  PSF star, which was observed shortly before or after the science
  observations. We would conclude that the narrow or broad component
  emission is truly extended if there is an excess in emission of the
  surface brightness profile compared to the bona-fide point like
  source represented by the PSF star.

  For determining the errors on our measurements, we created 100 mock
  data cubes by adding Gaussian random noise to the data based on the
  standard deviation of the spectra extracted from an object free
  region. The fitting procedure described earlier was repeated for
  each of these mock cubes (without any constraints on the parameters)
  and the error associated is the standard deviation of these 100
  measurements.
	
  The outflow velocities were estimated from the [\ion{O}{iii}] $\lambda$5007
  in the integrated spectrum of the objects using different
  prescriptions adopted previously in the literature. Due to the low S/N
  in each spaxel despite the re-binning procedure, we did not create
  velocity maps. We used non-parametric analysis of the line profile
  and refer the reader to \cite{zakamska14, liu13, rupke13} and
  \cite{perna15} among others for a detailed description of this
  procedure. The advantages of using a non-parametric measurement is
that it does not depend on the details of the fitting procedure and
the properties of the individual Gaussian components in the
model. After subtracting the continuum and reproducing the residual
line profile, the velocities at different percentiles are evaluated
using the cumulative flux function
$F(v) = \int_{-\infty}^{v} F_{v}(v^{\prime})dv^{\prime}$, where
$F_{v}$ is the line profile in the velocity space. The zero velocity
is taken to be at the peak of the line profile, which is the
redshifted wavelength of [\ion{O}{iii}] $\lambda$5007. The velocity at $x$
percentile, $\mathrm{v_{x}}$ is defined as the velocity at which the
cumulative flux function reaches $x$ percent of the overall flux of
the asymmetric line profile. Throughout this work, we have
  calculated $\mathrm{v_{10}}$ and $\mathrm{w_{80} = v_{90} - v_{10}}$
  for the entire [\ion{O}{iii}] $\lambda$5007 profile on the integrated spectra
  of the objects i.e. the velocities at the 10th percentile and the
  width containing 80\% of the total flux respectively and
  $\mathrm{v_{10}}$ on the broad-only Gaussian component as estimates
  of outflow velocity.

  These diagnostics have been used in the literature to estimate the
  outflow velocities and we will compare these values for our targets
  to estimate the uncertainties affecting such measurements (see
  Sec. \ref{sec_discussion}). Finally, in the kinematic analysis, we
  assume that the highest velocity is reached in the outermost region
  of the outflow.

\section{Results}    \label{sec_results}

In the following subsections, we will present the results of the data
analysis described above for the individual targets. We refer the
reader to Table \ref{all_linefit_parameters} for the results of line
fitting for the integrated spectra of each object.

\subsection{XID178 and XID5627}   \label{sec_results_XID178_XID5627}

\begin{figure*}
\centering
\begin{subfigure}{0.7\textwidth}
\centering
\includegraphics[width=11.5cm, height=8cm]{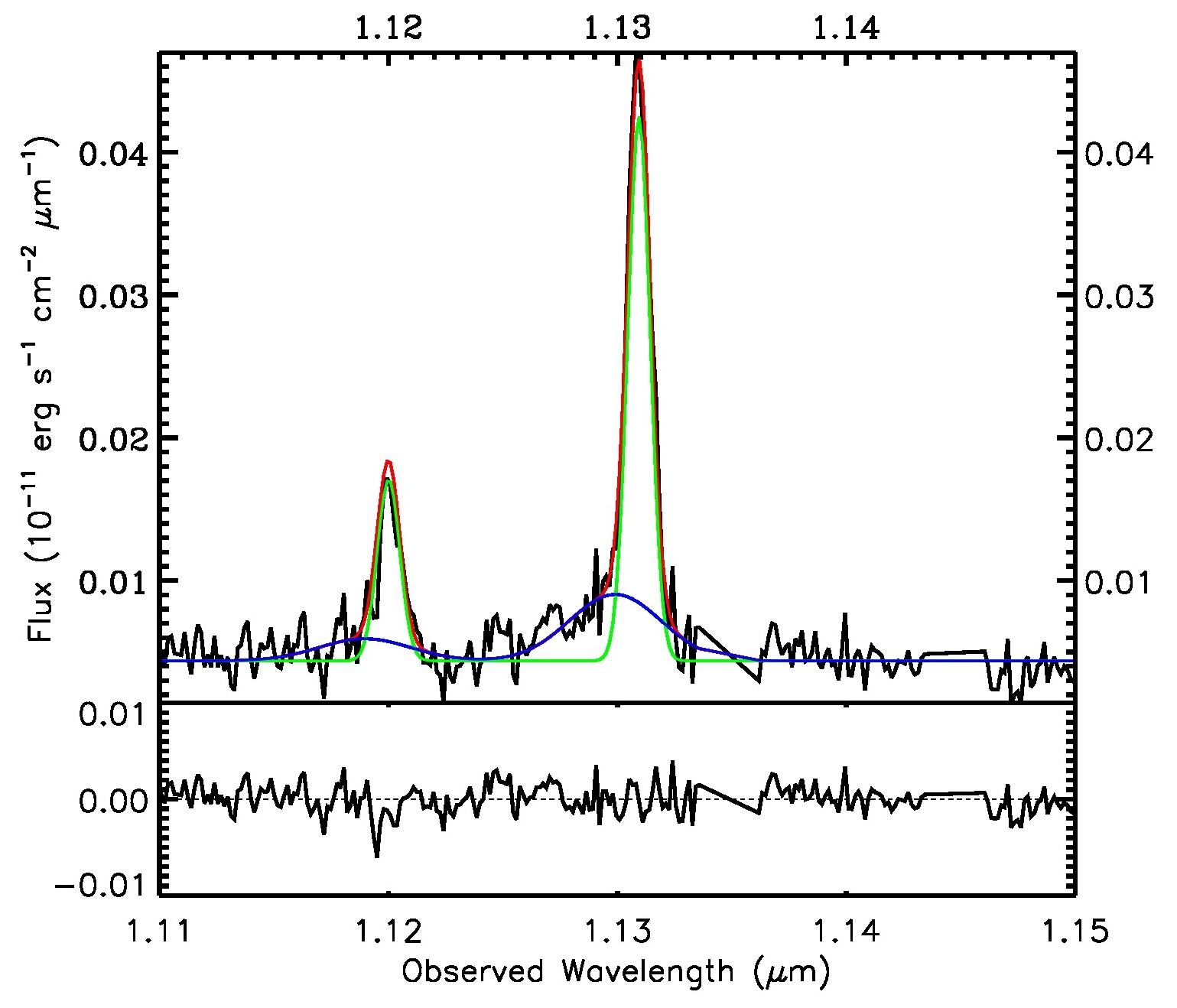}
\end{subfigure}
\begin{subfigure}{0.7\textwidth}
\centering
\includegraphics[width=11cm, height=5.8cm]{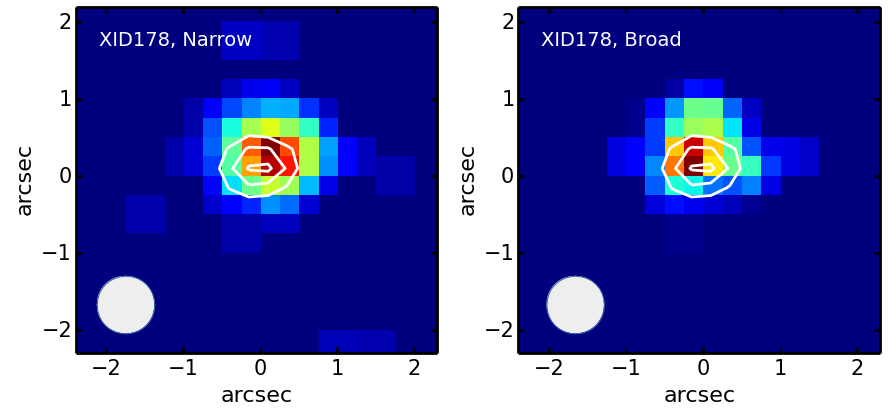}
\end{subfigure}
\begin{subfigure}{0.7\textwidth}
\centering
\includegraphics[width=8.5cm, height=6cm]{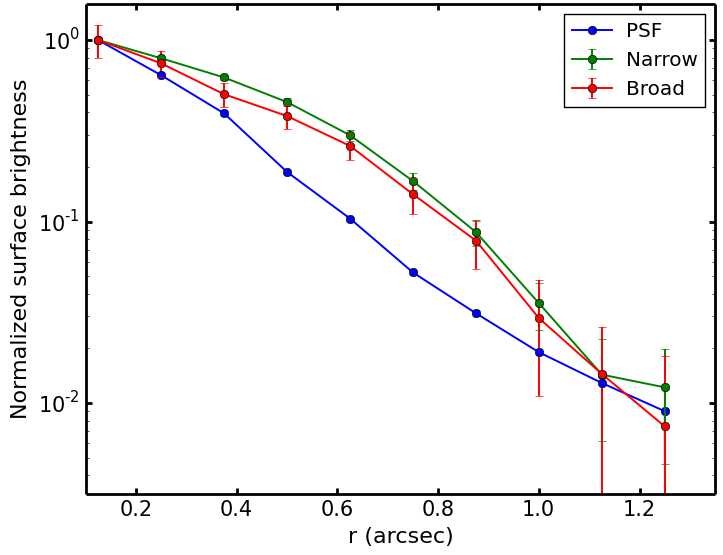}
\end{subfigure}
\caption{{\it Upper Panel:} The J band spectrum of XID178 around
    the [\ion{O}{iii}] $\lambda$4959, 5007 lines extracted from a region of
    1.25"$\times$1.25" centered at the QSO position. The observed
    spectrum is shown in black, the red curve shows the full fit, the
    green curve shows the narrow line component while the blue curve
    shows the broad line component. The residuals of the fit are shown
    in the smaller panel below the integrated spectrum. {\it Middle
      panel:} maps of the narrow (left) and broad (right) component in
    the [\ion{O}{iii}] $\lambda$5007 profile for XID178. Each pixel corresponds
    to a spatial scale of 0.25" which at the redshift of this object
    is equivalent to 2.15 kpc.The white contours in these maps
    represent the respective continuum emission of the object at
    levels 50\%, 75\% and 95\%. North is up and East is left. The maps
    show a one pixel offset between the narrow and the broad
    components. The gray circle on the lower left shows the size of PSF.~ {\it Lower panel:} Surface brightness profiles of the
    individual Gaussian components in XID178 derived from the flux
    maps in the middle panel. Green curve denotes the narrow
    component, red curve the broad component and the blue curve is the
    PSF of the observations. Both components show an
    extension up to a distance of 1". \label{178}}
\end{figure*}

\begin{figure*}
\centering
\begin{subfigure}{0.7\textwidth}
\centering
\includegraphics[width=11.5cm, height=8cm]{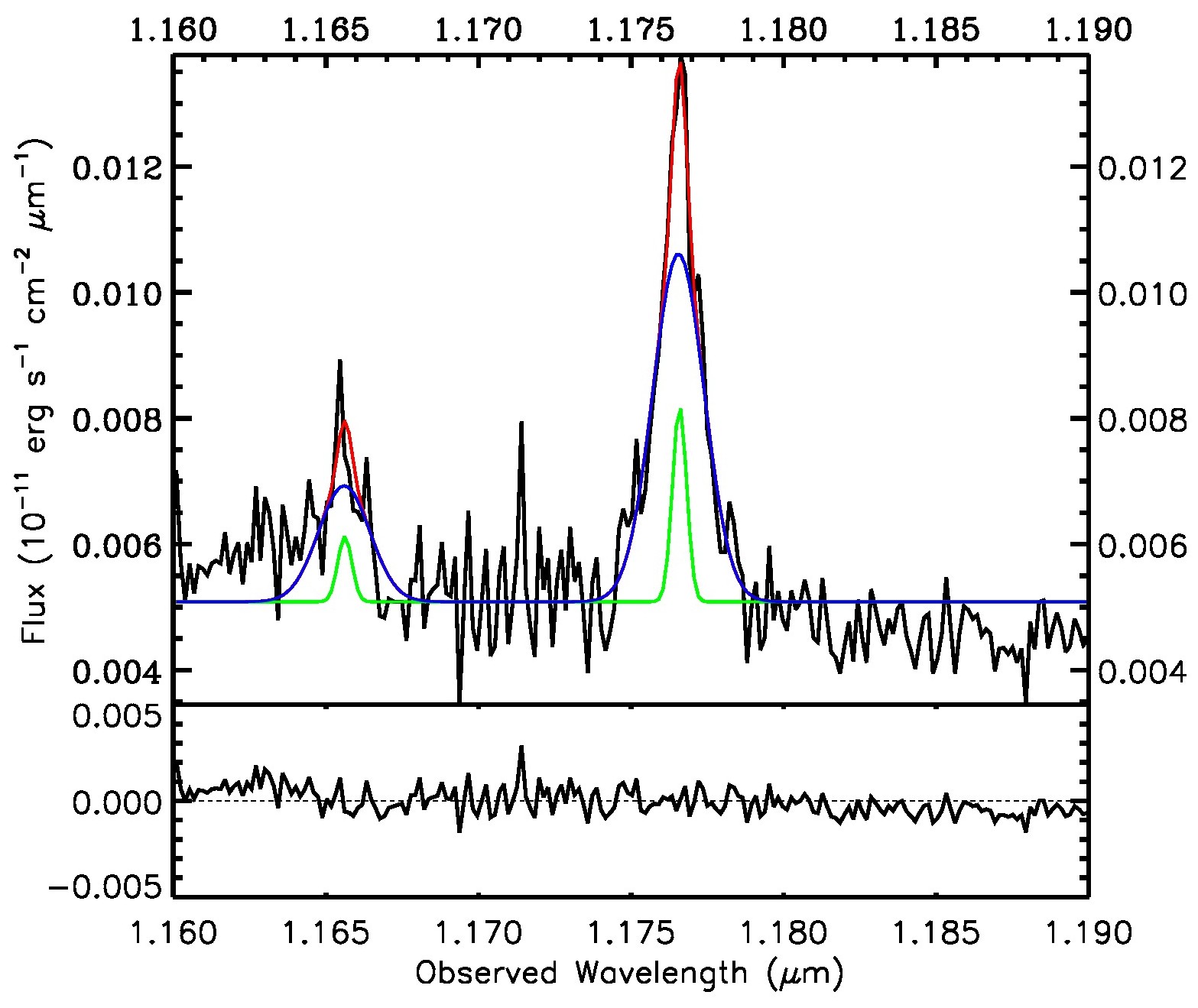}
\end{subfigure}
\begin{subfigure}{0.7\textwidth}
\centering
\includegraphics[width=11cm, height=5.8cm]{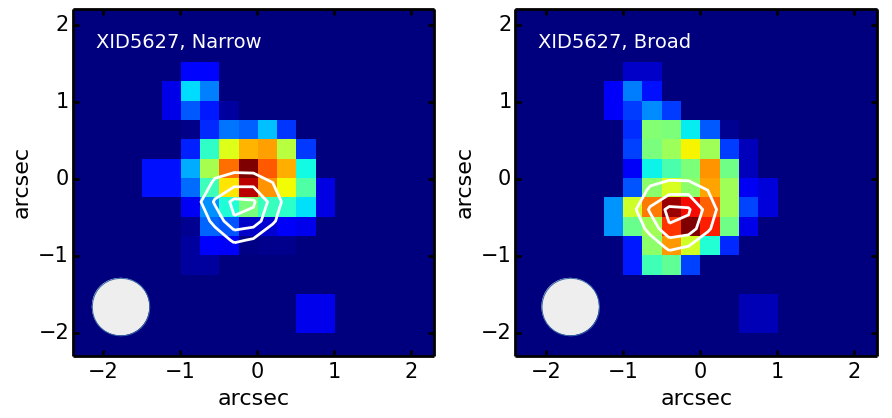}
\end{subfigure}
\begin{subfigure}{0.7\textwidth}
\centering
\includegraphics[width=8.5cm, height=6cm]{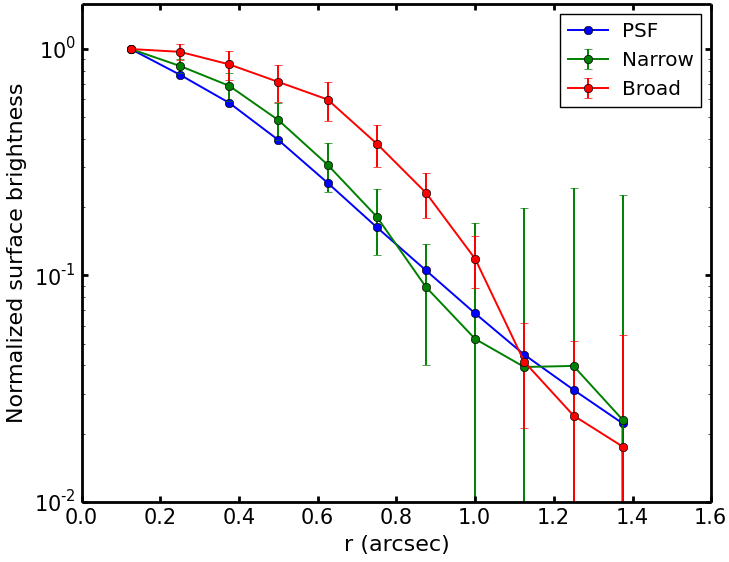}
\end{subfigure}
\caption{{\it Upper Panel:} The J band spectrum of XID5627 around
    the [\ion{O}{iii}] $\lambda$4959, 5007 lines extracted from a region of
    1"$\times$1" centered at the QSO position. The color coding for
    the plot is the same as that of XID178 in Fig. \ref{178}. The
    residuals of the fit are shown in the smaller panel below the
    integrated spectrum. {\it Middle panel:} maps of the narrow (left)
    and broad (right) component in the [\ion{O}{iii}] $\lambda$5007 profile for
    XID5627. Each pixel corresponds to a spatial scale of 0.25" which
    at the redshift of this object is equivalent to 2.16 kpc.The white
    contours in these maps represent the respective continuum emission
    of the object at levels 50\%, 75\% and 95\%. North is up and East
    is left. The maps show a significant offset between the spatial
    locations of narrow and broad emission.The gray circle on the lower left shows the size of PSF.~ {\it Lower panel:}
    Surface brightness profiles of the individual Gaussian components
    in XID5627 derived from the flux maps in the middle panel. Green
    curve denotes the narrow component, red curve the broad component
    and the blue curve is the PSF of the observations. 
    Only the broad component shows an extension. The narrow component
    is point-like since it is consistent with the profile of the PSF
    star. \label{5627}}
\end{figure*}

XID178 and XID5627 are the candidates in our sample \citep[apart from
XID2028 which was published in][]{cresci15} which show definite
evidence for the presence of outflow from the [\ion{O}{iii}] $\lambda$5007
profile. At a redshift of 1.253 and 1.337, both targets were observed
using the J grating of SINFONI to sample the [\ion{O}{iii}]  lines at $\lambda$
= 4959 $\AA$ and 5007 $\AA$. The total exposure time on source was 3.5
hours for each object achieving a S/N ratio of 34 (XID178) and 11
(XID5627) on the integrated spectrum for the [\ion{O}{iii}] $\lambda$5007
line. Due to the low S/N of the [OIII]$\lambda$4959 profile in both 
targets, the analysis has been restricted to [\ion{O}{iii}] $\lambda$5007
only. Table \ref{all_linefit_parameters} lists the results of the line
fitting procedure described in
Sec. \ref{sec_data_analysis}. Figs. \ref{178} and \ref{5627} show the integrated spectrum, the flux maps corresponding to narrow and broad Gaussian components and the surface brightness profiles of XID178 and XID5627 respectively.
% observed wavelength for XID5627 1.17662

In order to obtain the flux maps and therefore determine the spatial
distribution of the outflow, we re-binned the reduced cube by
clustering a 2$\times$2 group of pixels into a single pixel for both
the targets. This increases the S/N and allows for a better line
fitting across the spaxels. As a result, the re-binning procedure
increases the spatial scale of the pixel by a factor of 2 to 2.15 kpc
and 2.16 kpc for XID178 and XID5627 respectively.

The [\ion{O}{iii}] $\lambda$5007 profile in the integrated spectrum of XID178
in Fig. \ref{178}, top panel shows a clear blue wing which traces outflowing
gas moving towards us. The velocity offset between the narrow and the
broad line Gaussian component is about -275 km/s. The blue wing is not
present in the [\ion{O}{iii}] $\lambda$4959 profile, but we notice that its
location coincides with a significant telluric absorption line as can
be seen in the residual spectra. Moreover, as mentioned earlier, the
[\ion{O}{iii}] $\lambda$4959 profile has too low a S/N to draw any conclusions
about the presence of an outflow. Unfortunately, the presence of a red
wing in [\ion{O}{iii}] $\lambda$5007 cannot be verified since this region of
the spectrum is also affected by telluric absorption features as well.

The location of the outflow is apparent from the spatial offset
  between the narrow and the broad component maps in
  Fig. \ref{178}, middle panels. The white contours
  in the figure represent the location of the continuum emission. 
  The broad emission appears slightly shifted towards east
  with respect to the continuum as well as the narrow profile. The surface brightness profiles of these individual
  components in Fig. \ref{178}, bottom panel confirm that
  they are extended since there is an excess in the
  [\ion{O}{iii}] $\lambda$5007 emission compared to the PSF profile up to 1"
  ($\sim$ 8.6 kpc).

The integrated spectrum of XID5627 around the [\ion{O}{iii}] $\lambda$4959,5007
is shown in Fig. \ref{5627}, top panel. The [\ion{O}{iii}] $\lambda$5007
line profile is nearly symmetric, though two Gaussian components were
required to reproduce the overall extended line profile, with a
velocity offset of about -9 km/s between the narrow and the broad
Gaussian components. The line profile thus suggests that the ionized
gas might be moving both towards and away from the observer.

The middle panels of Fig. \ref{5627} show the map tracing the narrow and the broad components of [\ion{O}{iii}] $\lambda$5007 profile in XID5627. There is a clear
  offset between the spatial locations of the narrow and broad
  component emissions. In addition, the narrow component is
  point-like since it is consistent with the PSF profile, while the broad
  component is extended up to a distance of 1" ($\sim$ 8.40 kpc) from
  the center (Fig. \ref{5627}, bottom panel).

  The outflow velocities have been estimated from the integrated
  spectrum as described in Sec. \ref{sec_data_analysis}. The values
  from the different prescriptions are tabulated in Table
  \ref{outflow_velocity_table}. The error values are 1$\sigma$ errors
  computed by creating 100 mock spectra as described in
  Sec. \ref{sec_data_analysis}. The velocity values cover the range
  $\sim 490 - 730$ and $\sim 250 - 520$ km/s for XID178 and XID5627
  respectively. Clearly, these ranges of velocities will imply a range
  in the outflow properties calculated in
  Sec. \ref{sec_outflow_properties}, as discussed in
  Sec. \ref{sec_discussion}.

\begin{table}
\centering                          
\begin{tabular}{c c c c c}        
\hline\hline                 
 & XID178 & XID5627 & XID5330 & XID54204 \\    
\hline                       
$\mathrm{v_{10}^{a}}$ & -493 $\pm$ 27  & -247 $\pm$ 21 & -258 $\pm$ 21 & -1265 $\pm$ 32\\
$\mathrm{w_{80}^{a}}$ &  ~730 $\pm$ 60 &  ~517 $\pm$ 29 &  ~659 $\pm$ 21 &  ~1879 $\pm$ 135\\
$\mathrm{v_{10}^{b}}$ & -665 $\pm$ 31 & -265 $\pm$ 28 & -- & -963 $\pm$ 584\\
\hline                                 
\end{tabular}
\caption{The velocity values derived from the integrated spectrum. $\mathrm{v_{10}}$ denotes the 10th percentile velocity value, $\mathrm{w_{80}}$ denotes the velocity width containing 80\% of the flux. \label{outflow_velocity_table}}
\tablefoot{$^{a}$ Derived from the integrated spectra. $^{b}$ for broad component only}
\end{table}

\subsection{XID5330}    \label{sec_results_XID5330}

\begin{figure*}
\centering
\begin{subfigure}{0.45\textwidth}
\centering
\includegraphics[width=9cm, height=7cm]{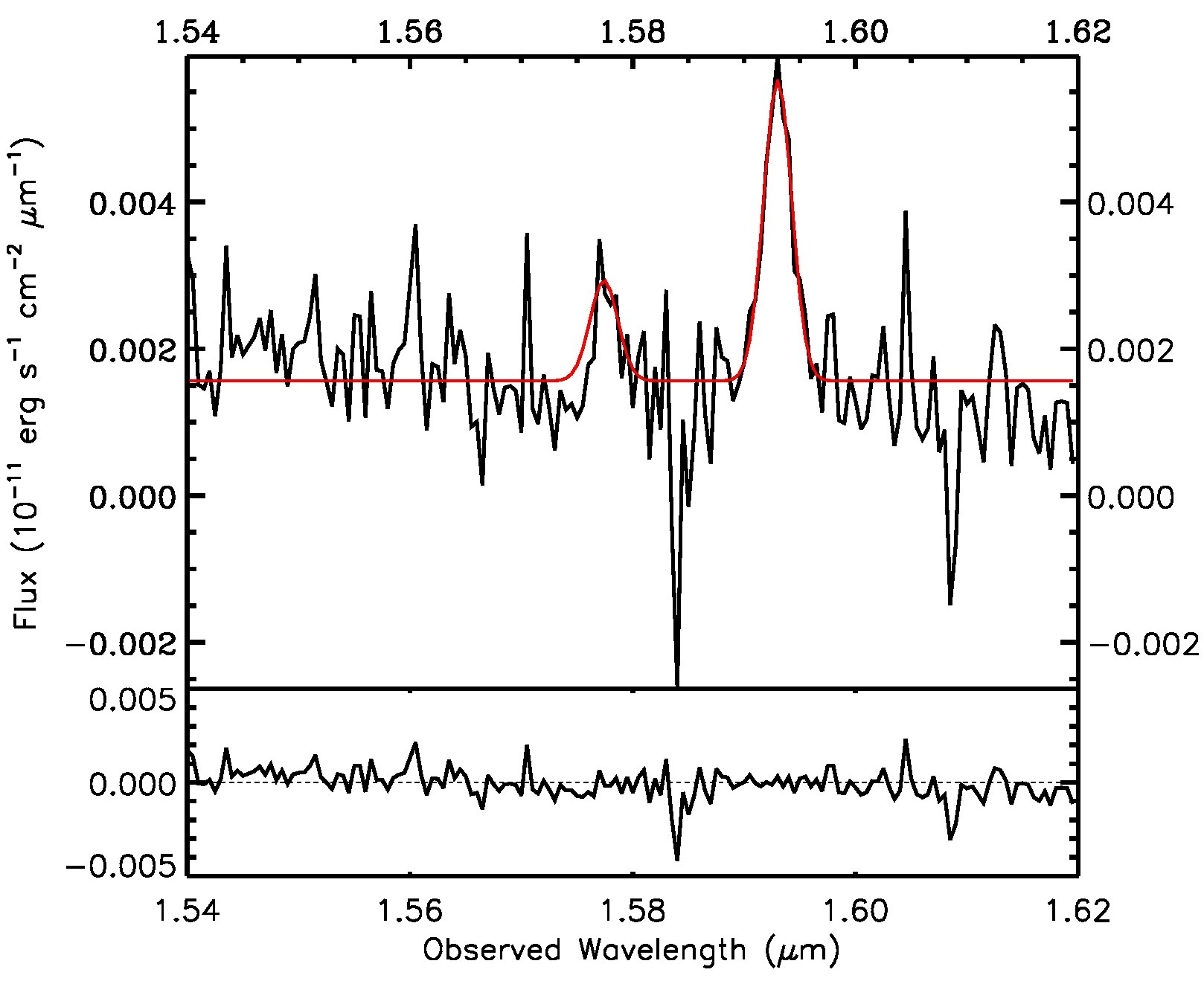}
\end{subfigure}
\begin{subfigure}{0.45\textwidth}
\centering
\includegraphics[width=7cm, height=7cm]{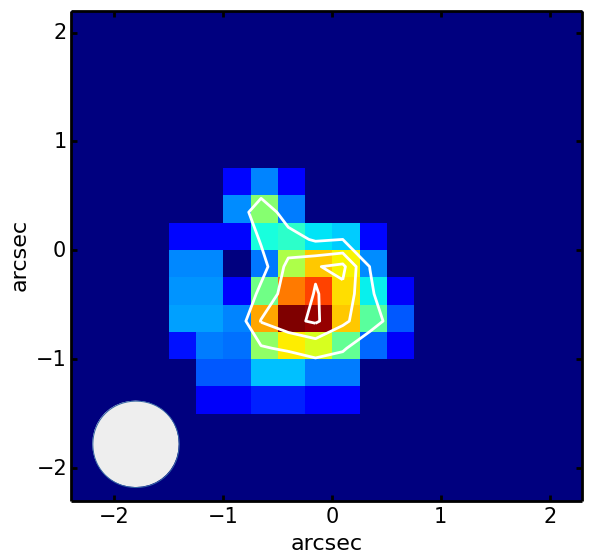}
\end{subfigure}
\begin{subfigure}{0.7\textwidth}
\centering
\includegraphics[width=8.5cm, height=6cm]{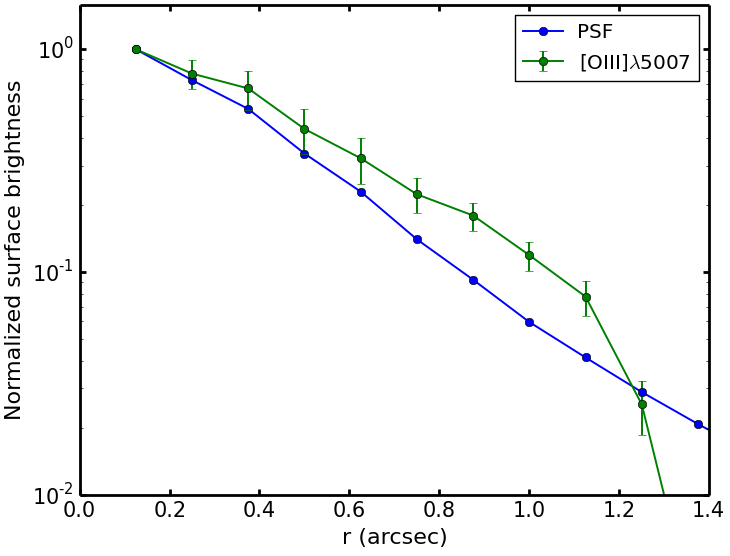}
\end{subfigure}
\caption{{\it Top left:}The H band spectra of XID5330 around
  [\ion{O}{iii}] $\lambda$4959,5007 lines extracted from a region of
  1"$\times$1" centered at the QSO position. The observed spectrum in
  shown in black and the full fit model is shown in red. Due to low
  S/N on the spectra, only a single Gaussian component was used for
  line fitting. The lower panel shows the residuals of the line
  fit. {\it Top right:} Map tracing the Gaussian component in
  [\ion{O}{iii}] $\lambda$5007 profile of XID5330. The white contours show the
  continuum levels at 50\%, 75\% and 95\%. Each pixel corresponds to a
  spatial scale of 0.25" ($\sim$2.12 kpc). North is up and east is
  towards left. The gray circle on the lower left shows the size of PSF.~{\it Bottom panel:} The surface brightness profile of
  the narrow component of the map in the top right panel compared to
  the PSF profile. The [\ion{O}{iii}] $\lambda$5007 emission is extended up to
  1.1" ($\sim$ 9.5 kpc). \label{5330_h}}
\end{figure*}

\begin{figure*}
\centering
\includegraphics[scale=0.15]{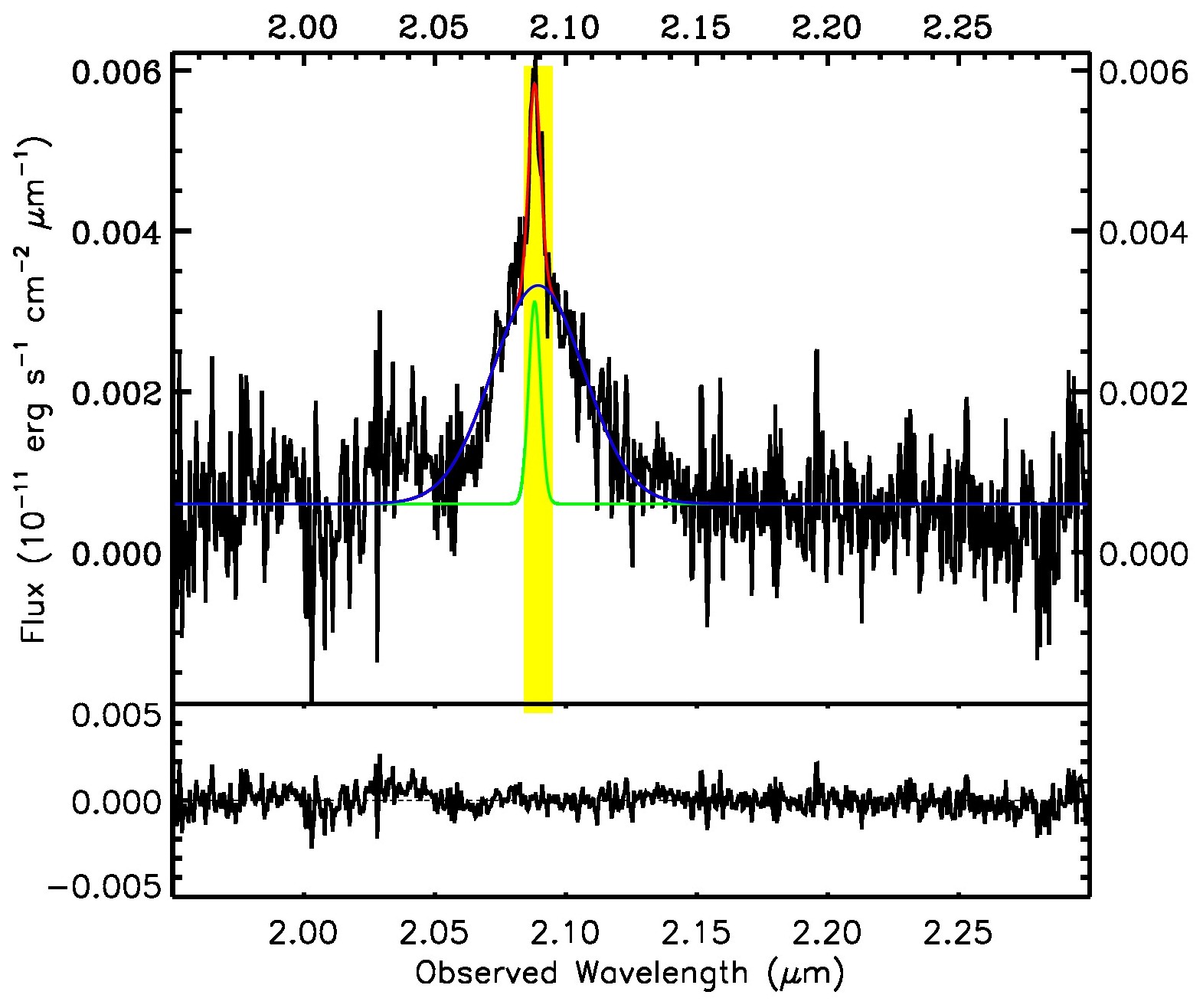}
\caption{The K band spectra of XID5330 around
  H$\alpha$ line extracted from a region of 1"$\times$1" centered at
  the QSO position. There are two distinct components of overall
  H$\alpha$ line profile: a narrow component, shown in green and a
  broad component, shown in blue. The overall fit is shown in red. The
  lower panel shows the residual of this line fit. \label{5330_k}}
\end{figure*}

XID5330 has the shortest integration time among the targets of our
program. The spectral resolution is lower compared to the rest of the
sample since the observations were taken in the SINFONI H+K band. At a
redshift of 2.169, the [\ion{O}{iii}]  line falls in the H band at a wavelength
of 1.593 $\mu$m and we also detect H$\alpha$ line at 2.088 $\mu$m in
the K band. In $\sim$50 minutes, we were able to reach a S/N of 5 for
[\ion{O}{iii}] $\lambda$5007 and 8.3 for H$\alpha$ in the integrated spectrum.

The integrated H band spectrum around the [\ion{O}{iii}] $\lambda$4959,5007
lines is shown in Fig. \ref{5330_h}, top left panel. We have a low S/N
ratio for the [\ion{O}{iii}] $\lambda$5007 profile and hence only one Gaussian
has been used for line fitting. Fig. \ref{5330_h}, top right panel shows the map tracing the narrow
  component which shows an extension of
  $\sim$1.1" , which translates to a spatial scale of 9.5 kpc from the
  centre, when compared to the PSF profile
  (Fig. \ref{5330_h}, bottom panel). The absence of a broad Gaussian
  component coupled with low spectral resolution and low S/N on the
  spectrum does not allow us to compute mass outflow rates and the
  associated kinetic power as done for other objects in the sample.

The K band spectrum around the H$\alpha$ line profile, shows a clear
distinction between the narrow and broad H$\alpha$ components in the
spectrum (see Fig. \ref{5330_k}, left panel). The line fitting
parameters for H$\alpha$ are given in Table
\ref{5330_ha_fit_parameters}. The broad component of H$\alpha$ has a
velocity dispersion of $\sim$ 6000 km/s and is clearly AGN
related. The velocity offset between the narrow and the broad Gaussian
components is about +190 km/s (the broad component is redshifted with
respect to the narrow component). The low S/N of the spectra does not
allow us to disentangle contributions (if any) to the flux from
[NII]$\lambda$6549,85 lines which are expected to lie at 2.0837$\mu$m
and 2.0951$\mu$m respectively i.e. on either side of the narrow
H$\alpha$ component.

\begin{table}
\centering                          
\begin{tabular}{c c}        
\hline\hline                 
Parameter & Value \\    
\hline                       
   $\mathrm{\lambda_{narrow}}$  &  2.0881 $\pm$ 0.0005 $\mu$m\\     
   $\mathrm{v_{narrow}}$        &  770 $\pm$ 282 km/s      \\
   $\mathrm{\lambda_{broad}}$  &  2.0894 $\pm$ 0.0005 $\mu$m\\
   $\mathrm{v_{broad}}$         &  5950 $\pm$ 808 km/s     \\
\hline                                 
\end{tabular}
\caption{Line fitting parameters for K band spectra of XID5330. Two components, narrow and broad Gaussian were used to fit the entire H$\alpha$ profile. $\mathrm{\lambda_{narrow}}$ = central wavelength of the narrow Gaussian component;  $\mathrm{v_{narrow}}$ = velocity corresponding to the FWHM of the broad Gaussian; $\mathrm{\lambda_{narrow}}$= central wavelength of the broad Gaussian component and $\mathrm{v_{broad}}$ = velocity corresponding to the FWHM of the broad Gaussian component. The wavelength of the broad component is redshifted by 190 km/s with respect to the narrow component.   \label{5330_ha_fit_parameters}}
\end{table}

The presence of the broad H$\alpha$ line gives us the possibility
to estimate and verify the mass of the black hole.  We use the
\cite{greene05} formalism wherein the black hole mass from the
H$\alpha$ line is given by:

\begin{equation}
\label{eq_bhmass}
m_{bh}^{H\alpha} = 2.0 \times 10^{6} \left(
\frac{L_{H\alpha}}{10^{42} ~\mathrm{erg ~s^{-1}}}
\right)^{0.55}\left(
\frac{FWHM_{H\alpha}}{1000 ~\mathrm{km ~s^{-1}}}
\right)^{2.06} M_{\odot}
\end{equation}

We consider only the broad line component of the flux of the H$\alpha$
profile for the estimation of luminosity, which comes out to be
$\mathrm{2.7\pm 1.0 \times 10^{43}}$ erg/s. Using the values from
Table \ref{5330_ha_fit_parameters} and the luminosity of the broad
component, we arrive at a black hole mass of 8.9$\pm$0.2 $\times$
10$^{8}$ M$_{\odot}$. This is a factor of $\sim$5 larger than the
value given in Table \ref{sample_selection}, which is based on broad
MgII line. This discrepancy might be due to the low S/N in both
optical as well as H+K band data. Moreover, we are unable to
disentangle the contribution due to [NII]$\lambda$6549,85 on the
entire H$\alpha$ profile which, if present, could bring down our mass
estimate.

\subsection{XID54204}    \label{sec_results_XID54204}

\begin{figure*}
\centering
\begin{subfigure}{0.45\textwidth}
\centering
\includegraphics[width=9cm, height=6cm]{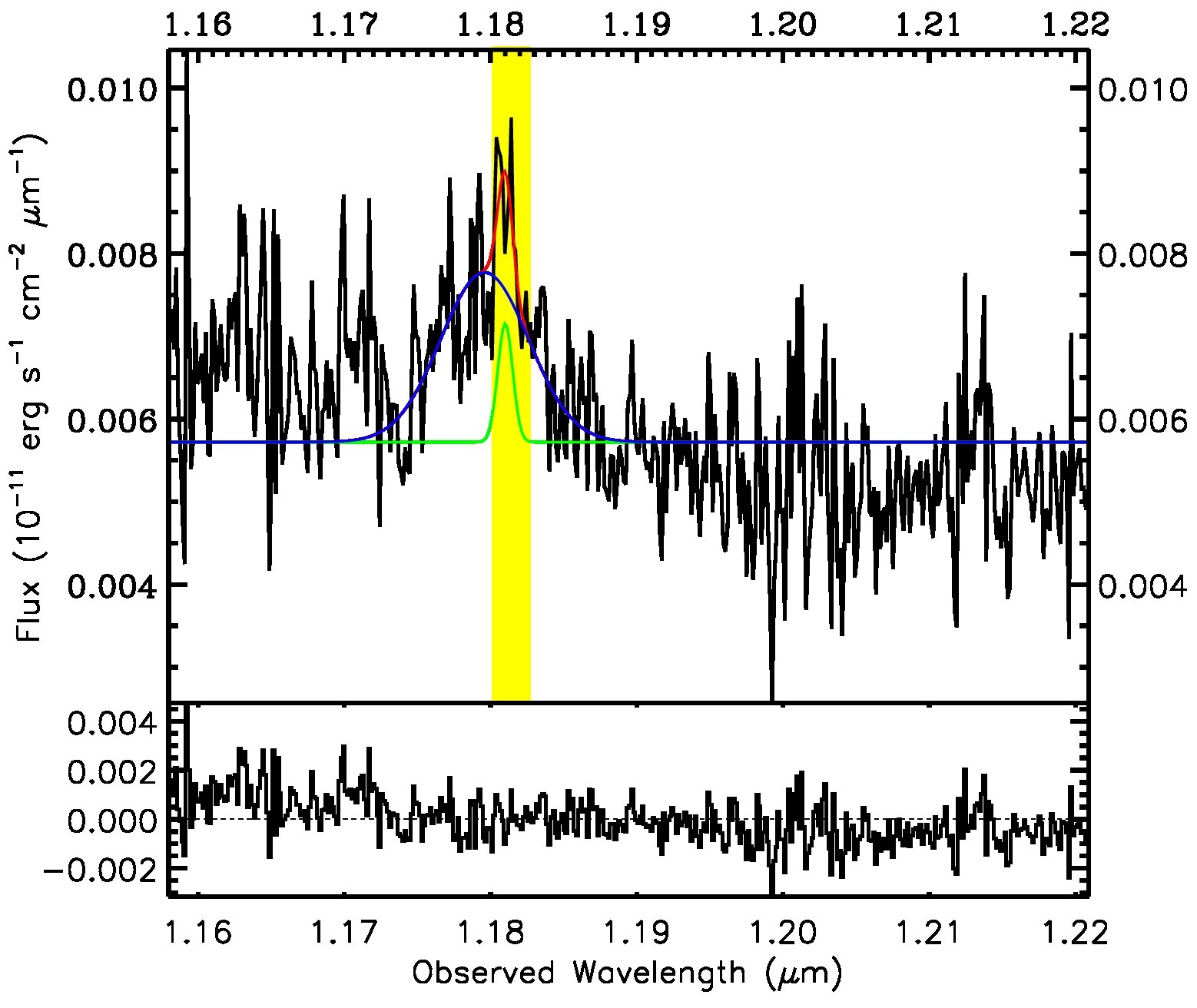}
\end{subfigure}
\begin{subfigure}{0.45\textwidth}
\centering
\includegraphics[width=6cm, height=6cm]{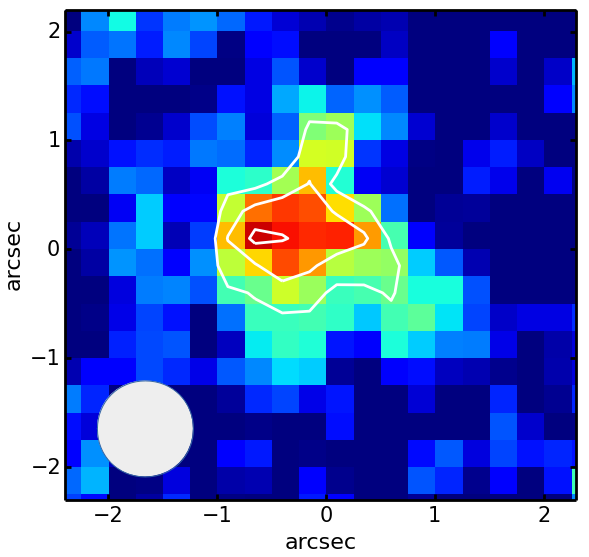}
\end{subfigure}
\caption{{\it Left panel:} The J band spectrum of XID54204 around
  [\ion{O}{iii}] $\lambda$5007 line extracted from a region of 1"$\times$1"
  centred at the QSO position. The color codes for different part of
  the spectra are defined in the same way as Fig. \ref{178}, top panel. {\it Right Panel:} Continuum collapsed channel map around the narrow
  component of [\ion{O}{iii}] $\lambda$5007 line (yellow region in the left panel, 1.180-1.182 $\mu$m). Each pixel corresponds to a spatial scale of
  0.25" which at the redshift of this object is equivalent to $\sim$2.16
  kpc. The white contours indicate the peak of the [\ion{O}{iii}] $\lambda$5007
  emission at levels 50\%, 75\% and 95\%. North is up and East is
  towards left. The gray circle on the lower left shows the size of PSF.~\label{54204_flux}}
\end{figure*}

XID54204 has a very faint [\ion{O}{iii}] $\lambda$5007 signal despite an
integration time of 3.5 hours on the target. The [\ion{O}{iii}] $\lambda$5007
line is detected at a wavelength of 1.1796 $\mu$m consistent with the
object being at z$=$1.356. A S/N of 4.4 could be reached in the
integrated spectrum of the [\ion{O}{iii}] $\lambda$5007 emission line.

Although faint, the [\ion{O}{iii}] $\lambda$5007 profile in the integrated
spectrum is quite broad, so a single Gaussian is not able to provide a
good fit of the line profile. Therefore we added a second component
and the resulting integrated spectrum is given in
Fig. \ref{54204_flux}, left panel. The integrated spectrum indicates
the presence of a blue wing in the [\ion{O}{iii}] $\lambda$5007 profile, hence
the outflow direction is towards the observer. The velocity offset
between the narrow and the (blue-shifted) broad Gaussian component is
$\sim$360 km/s.

Fig. \ref{54204_flux}, right panel shows the continuum subtracted
  [\ion{O}{iii}] $\lambda$5007 line core map, confirming [\ion{O}{iii}]  emission from this object. The line core map was obtained by integrating the continuum collapsed SINFONI data cube between 1.180-1.182 $\mu$m, shown by the yellow region in the integrated spectrum in Fig. \ref{54204_flux}. Due to low S/N in each spaxel, one cannot do line fitting across the FOV and consequently we could not obtain surface brightness profiles as we did for other objects. In this case, one can
  only put an upper limit of the spatial extension of the outflow from
  the FWHM of the PSF, which is 0.8" ($\sim$ 6.9 kpc). The outflow
  velocity from the integrated spectrum have been estimated in the
  range $\sim$950-1900 km/s (see Table \ref{outflow_velocity_table}).

\section{Outflow properties}    \label{sec_outflow_properties}

The determination of the properties of the outflow such as its mass is
complicated by the fact that the outflow has a multi-phase
nature. While we are tracing the outflow in its warm ionized phase
using the [\ion{O}{iii}] $\lambda$5007 line, a significant fraction of the
outflows are believed to be in the cold molecular gas phase
\citep{veilleux13,cicone14, feruglio15, brusa15, nyland13}. There are
many papers in the literature which discuss the estimation of the mass
of the ionized outflows from either H$\beta$ or [\ion{O}{iii}] $\lambda$5007
\citep[e.g.,][]{holt11,cano-diaz12, liu13, harrison14}. Due to the limited information available to us (e.g., 
lack of Balmer lines), we derive the outflow gas mass from the [\ion{O}{iii}] $\lambda$5007 line based on \cite{cano-diaz12}. [\ion{O}{iii}]  lines are very sensitive to temperature and ionization but poor tracers of mass, hence the current analysis only provides (at best) an order-of-magnitude estimate of the outflow properties.

Below, we summarize the essential formulas used to
arrive at the outflow properties and the assumptions associated with
them. The mass of the ionized outflowing gas is given by:

\begin{equation}
\label{eq_outflowmass}
M_{ion}^{out} \approx 5.33 \times 10^{8} \left(
\frac{C}{10^{[O/H] - [O/H]_{\odot}}}
\right)
\left(
\frac{L[\ion{O}{iii}] }{10^{44} ~\mathrm{erg ~s^{-1}}}
\right)\left<
\frac{n_{e}}{100 ~\mathrm{cm^{-3}}}
\right>^{-1} M_{\odot}
\end{equation}
\noindent
where C (= <n$_{e}$>$^{2}$/<n$_{e}^{2}$>) $\sim$ 1 based on the
hypothesis that all the ionizing gas clouds have the same density,
L[\ion{O}{iii}]  is the luminosity of the [\ion{O}{iii}] $\lambda$5007 line tracing the
outflow, n$_{e}$ is the electron density of the outflowing gas and
10$^{[O/H] - [O/H]_{\odot}}$ represents the oxygen abundance in Solar
units. The luminosity of [\ion{O}{iii}] $\lambda$5007 is derived from the flux
of the broad line component in the integrated spectrum, when
present. In the absence of a broad component, mass estimate of the
outflowing gas could not be obtained. The normalization factor in
Eq. \ref{eq_outflowmass} is dependent on the emissivity of the ionized
gas which is further dependent on electron density and the temperature
of the gas. The emissivity calculations were done using PyNeb
\citep{luridiana15}. The following assumptions go into
Eq. \ref{eq_outflowmass}:
\begin{itemize}
\item Most of the oxygen in the outflow is in its ionized form,
  O$^{+2}$ i.e. n(O$^{+2}$) $\approx$ n(O).

\item The number density of atomic helium is 10\% that of atomic
  hydrogen i.e. n(He) $\approx$ 0.1 n(H). This is based on the "cosmic
  composition" from abundance measurements in the Sun, meteorites and
  other disk stars in the Milky Way \citep{ferriere01}.

\item The electron density, n$_{e}$ is assumed to be equal to 100
  cm$^{-3}$. The electron density is usually measured using the line
  ratio of the [\ion{S}{ii}] $\lambda$6716,31 doublet \citep{peterson97}. As a
  recent example of this technique, \cite{perna15} estimate
  n$_{e}\approx$ 120 cm$^{-3}$ from the [\ion{S}{ii}]  doublet in the outflow
  component for a high S/N target XID5321 in the COSMOS field which is
  within the typical value estimates of electron density for the NLR
  ($\mathrm{\lesssim 50-1500 ~cm^{-3}}$). Previous works also use a
  value within this range when the electron density is not known. For
  example, \cite{cano-diaz12} use a value of 1000 cm$^{-3}$, while
  \cite{cresci15} and \cite{liu13} assume a value of 100 cm$^{-3}$ and
  \cite{harrison14} and \cite{carniani15} use a value of 500
  cm$^{-3}$. We have taken these differences into account while
  comparing the outflow rates we derive to the previous works in
  Sec. \ref{sec_discussion}. We also discuss the level of
  uncertainties due to these assumptions.

\item Emission lines from a single ion having different excitation
  potentials are highly temperature dependent which makes them
  suitable for electron temperature measurement in the NLR
  \citep{peterson97}. As mentioned before, the normalization factor in
  Eq. \ref{eq_outflowmass} depends on the emissivity of the ionized
  gas which is sensitive to temperature changes. Hence the
  normalization factor in Eq. \ref{eq_outflowmass} will change with
  the assumed temperature. Usually, [\ion{O}{iii}] $\lambda$4363,4959,5007 or
  [NII]$\lambda$5755,6548,6583 are the set of lines used for this
  purpose. Typical temperatures measured for the NLR using this method
  come out to be about 10,000-25,000 K. For our calculations, we
  assume a value of 10,000 K.

\item Since the current data do not allow us to determine oxygen abundances, solar metallicity values have been used.
\end{itemize} 

Due to the limited information available to us about the morphology of
the outflow, we assume a simple conical outflow model\footnote{It
  might actually be a bi-conical outflow morphology, but since we
  observe only one side of the outflow we only take into account this
  part in our calculations.} for our objects. The volume averaged
density of the outflowing gas, <$\mathrm{\rho_{out}}$> (not to be
confused with the gas density of individual clouds) is given by:

\begin{equation}
\label{eq_outflowdensity}
<\rho_{out}> = 3\frac{M_{out}}{\Omega\cdot R_{out}^{3}}
\end{equation}

where $\Omega$ is the solid angle subtended by the (bi)-conical
outflow and $\mathrm{R_{out}}$ is the extension of the outflow in the
cone. The mass outflow rate is then given by
$\mathrm{\dot{M}_{out} \approx <\rho_{out}>_{V}\cdot\Omega
  R_{out}^{2}\cdot v_{out}}$.
When combined with Eq. \ref{eq_outflowmass}, this gives:

\begin{align}
\label{eq_outflowmassrate}
\begin{split}
\dot{M}_{ion}^{out} \approx & ~1.64\times 10^3 \cdot \left(
\frac{1}{10^{[O/H] - [O/H]_{\odot}} R_{kpc}}
\right)
\left(
\frac{L[\ion{O}{iii}] }{10^{44} ~\mathrm{erg ~s^{-1}}}
\right)\\
& \left(
\frac{v}{1000 ~\mathrm{km ~s^{-1}}}
\right)
\left<
\frac{n_{e}}{100 ~\mathrm{cm^{-3}}}
\right>^{-1} M_{\odot}/yr
\end{split}
\end{align}

\noindent where v is the velocity of the outflowing gas (from
  Table \ref{outflow_velocity_table}) out to a radius
  $\mathrm{R_{kpc}}$ in units of kiloparsec. The kinetic
energy of the outflow due to the ionized component is simply
$\mathrm{1/2\dot{M}_{out}^{ion}v^{2}}$:

\begin{align}
\label{power}
\begin{split}
\dot{E}_{K}^{ion} =& ~5.17 \times 10^{44} \cdot 
\frac{1}{10^{[O/H] - [O/H]_{\odot}} R_{\mathrm{kpc}}} \left(
\frac{L[\ion{O}{iii}] }{10^{44} ~\mathrm{erg ~s^{-1}}} 
\right)\cdot \\
& \left(
\frac{v}{1000 ~\mathrm{km ~s^{-1}}}
\right)^{3}
 \left<
\frac{n_{e}}{100 ~\mathrm{cm^{-3}}}
\right>^{-1}  \mathrm{erg ~s^{-1}}
\end{split}
\end{align}

\noindent
If the distance of the outflowing region is known from the maps, we
can also compute the dynamical time scale of the outflow:
\begin{equation}
\label{dynamic_time}
t_d \approx R_{out}/v_{out}
\end{equation}
\noindent
where $\mathrm{v_{out}}$ is the outflow velocity inferred from the
kinematic analysis. The dynamical time scales could give us an idea
about the "on" phase of the outflow, which, whenever possible to
calculate, would be a direct observational constraint on the time
scales of the outflow driven by an AGN at high redshift. The outflow
properties mentioned in this section have been derived for every
object in our sample and are reported in Table
\ref{outflow_properties}. The Table indicates a range of these
  properties obtained using the velocity ranges in Table
  \ref{outflow_velocity_table}, keeping other assumptions such as the
  electron density, temperature, radius and metallicity constant. The
  errors on these properties have been ignored compared to the
  systematic uncertainties.

\begin{table*}
\centering          
\begin{tabular}{c c c c c c c c c c}     % 9 columns 
\hline\hline       
Object ID & $\mathrm{f_{[\ion{O}{iii}] }}$ & log($\mathrm{L_{[\ion{O}{iii}] }}$) & $\mathrm{M_{out}^{ion}}$ & $\mathrm{\dot{M}_{out}^{ion}}$ & $\mathrm{\dot{E}_{K}^{ion}}$ & $\mathrm{P_{SFR}}$\\ 
& (10$^{-16}$ erg/s/cm$^{-2}$) & (erg/s) & (10$^{7}$ M$_{\odot})$ & (M$_{\odot}$/yr) & (10$^{41}$ erg/s) & (10$^{43}$ erg/s)\\
(1) & (2) & (3) & (4) & (5) & (6) & (7)\\
\hline                    
   178    & 2.33 $\pm$ 0.16 & 42.35 $\pm$ 0.02  & 1.13 & 2.00-2.96 & 1.53-4.98   & 9.4  \\  
   5330*  & 1.29 $\pm$ 0.04 & 42.68 $\pm$ 0.01 & --   & --        & --          & 7.0 \\
   5627   & 1.65 $\pm$ 0.15 & 42.27 $\pm$ 0.02  & 0.94 & 0.83-1.74 & 0.16-1.46   & 11.0 \\
   54204  & 1.54 $\pm$ 0.84 & 42.26 $\pm$ 0.05  & 0.91 & 3.91-7.63 & 11.44-84.95 & 30.7 \\
   2028$^{\dagger}$  & 2.64 $\pm$ 0.09 & 42.66 $\pm$ 0.01 & 2.32 & 8.24 & 58.44 & 17.5\\
\hline                  
\end{tabular}
\caption{Outflow properties derived from the line fits and the formulas mentioned in Sec. \ref{sec_outflow_properties}. (1) The X-ray ID of the sample; (2) Flux of the broad Gaussian component in the [\ion{O}{iii}] $\lambda$5007 profile; (3) luminosity corresponding the flux in (2);  (4) is the outflow mass traced by the [\ion{O}{iii}] $\lambda$5007 line, (5) \& (6) are the range in 
mass outflow rates and the kinetic energy associated with the ionized component corresponding to the velocity values in Table \ref{outflow_velocity_table}, keeping the rest of the assumptions constant; (7) Power due to the star formation in the host galaxy. \label{outflow_properties}}
\tablefoot{*Mentioned values for XID5330 are for the overall line profile. \newline $^{\dagger}$Values reported are from [\ion{O}{iii}] $\lambda$5007 analysis from this work to be consistent while comparing with other objects.}
\end{table*}

\section{Discussion}   \label{sec_discussion}

\subsection{Outflow detection and selection efficiency}

\begin{figure}
\centering
\includegraphics[width=\hsize]{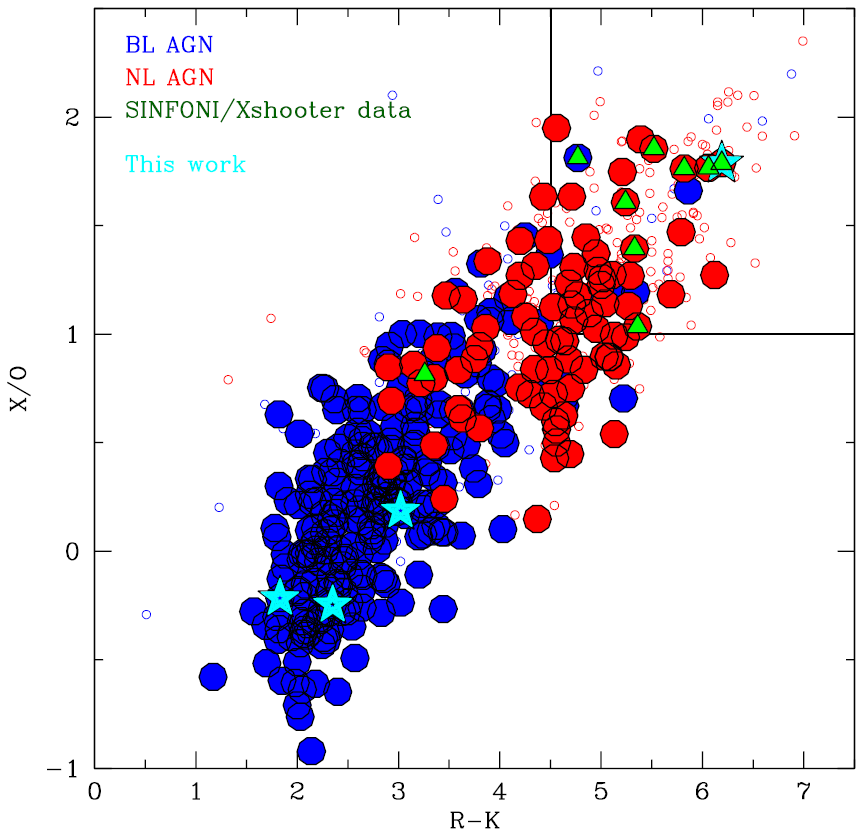}
\caption{X-ray to optical flux ratio (X/O) vs. R-K (Vega) color
  magnitude plot. Open circles represent all the data points from the
  XMM-COSMOS survey, red and blue filled circles represent the
  spectroscopically confirmed narrow line and broad line AGNs
  respectively. In green are the AGN with detected outflows published
  in \cite{brusa15a,perna15a}. Data points in starred cyan represent
  the objects from this paper. Clearly, with our sample, we are
  exploring a different region of the plane. \label{red_color}}
\end{figure}

At least three of the four targets (XID178, XID5627 and XID54204)
  presented in this paper show clear evidence of outflows. The S/N
ratio of XID54204 is very low so the spatial information of the
outflow could not be derived, although there is an indication of the
presence of a blue wing in its [\ion{O}{iii}] $\lambda$5007 profile in the
integrated spectrum. The low S/N of XID54204 could be simply 
due to the high air mass through which this
object was observed (see Sec. \ref{sec_observations}). Due to the low
integration time of XID5330 coupled with a lower spectral resolution
in the H+K band of the SINFONI grating, we were not able to reach the
S/N ratio limit for which we could observe the wings in the
[\ion{O}{iii}] $\lambda$5007 profile.  Including XID2028 \citep{cresci15} which
also satisfies the selection criterion of our sample, we can say that
Fig. \ref{forbidden_region} is an effective tool to select AGN in an
outflowing phase. Our study reinforces previous evidence for
absorption variability in the X-ray spectra of AGN close to the
forbidden region \citep{vasudevan13}. In order to populate this plane
with other known outflows, we have added in Fig. \ref{forbidden_region}
the sample from \cite{brusa15a} for which we have the hydrogen column
density and Eddington ratio estimates (red circles). We see that three
out of the four objects of that sample lie inside or very close to the
forbidden region which further points to the efficiency of this
selection procedure.

Previous studies on AGN outflows have been biased in pre-selected
targets with higher chances to be in an outflowing phase, similar to
what we did. As an example, the \cite{brusa15a} selection was based on
the fact that the QSOs in an outflowing phase are expected to be dusty
and reddened, the so-called "red QSOs". This is shown by the red
circles in Fig. \ref{red_color} which shows a plot between the X-ray
(2-10 keV) to optical flux ratio (X/O) and R-K (Vega) colors. Three of
our targets (marked by blue stars) lie in the blue QSO region (R-K
$\sim$ 2). XID54204 is not included in this figure as it is not
detected in the 2-10 keV band. XID2028 is the blue star in the red QSO
region. We are therefore exploring a completely different region of
the X/O vs. R-K plot and we find that the objects in the blue region
also show outflows. Clearly the next step would be to test these
selection criteria with a blind sample which covers the whole plane in
Fig. \ref{red_color} (and Fig. \ref{forbidden_region}).
  
The one dimensional spectra of three of the objects from our sample
(XID2028, XID5627 and XID54204 a.k.a XCOS2028, XCOS5627 and COS178
respectively) have also been presented in \cite{harrison15} as part
of the KMOS AGN survey at High Redshift (KASHz). Except for XID54204,
our one dimensional results (line wavelengths, widths and luminosity)
are compatible with those reported by \cite{harrison15}. The minor
differences might be due to the different apertures used for spectral
extraction from the IFU cube and the errors associated with the flux
calibration.  \cite{harrison15} do not detect any 
[\ion{O}{iii}] $\lambda$5007 in their analysis for XID54204 (COS178). We were
able to detect the presence of the faint [\ion{O}{iii}] $\lambda$5007 signal by
constructing a continuum subtracted channel map over the narrow
component of the line as given in Fig. \ref{54204_flux}.

\subsection{Source of the outflows}

In order to understand the nature of ionized outflows, we need to
estimate the total mass of the outflowing gas, the mass outflow rate
and the kinetic power driven by the AGN using the formulas given in
Sec. \ref{sec_outflow_properties}. As mentioned in
  Sec. \ref{outflow_properties}, we explore a range of outflow
  properties depending on the velocity measurements. Tight
correlations are expected between these quantities and the bolometric
luminosity of the AGN according to theoretical models which also
predict high velocity outflows from AGN \citep{king11,
  zubovas12}. These models predict outflow velocities in the range
$1000-1500$ km/s and mass outflow rates up to 4000 M$_{\odot}$/yr
which are extended to kiloparsec scales. XID2028 shows outflow
velocities up to 1500 km/s extended up to 13 kpc from the central
object with the total mass outflow rate expected to be at least 1000
M$_{\odot}$/yr \citep{cresci15}. Such high velocities could not be
sustained by star formation alone and a more powerful source such as
an AGN is required. This is also the case for XID54204 where we
  observe linewidths with velocity exceeding 1000 km/s. XID178 and
  XID5627 also show outflows which are spatially extended up to $\sim$
  8.6 kpc. Compared to XID2028 and XID54204, these objects have lower
  outflow velocities up to $\sim$700 km/s. However, these are
projected velocities and depending on the inclination of these
systems, the actual velocities may be higher than the reported
values. Nevertheless, these velocities are compatible with
star-formation processes.

\begin{figure*}
\centering
\includegraphics[scale=0.33]{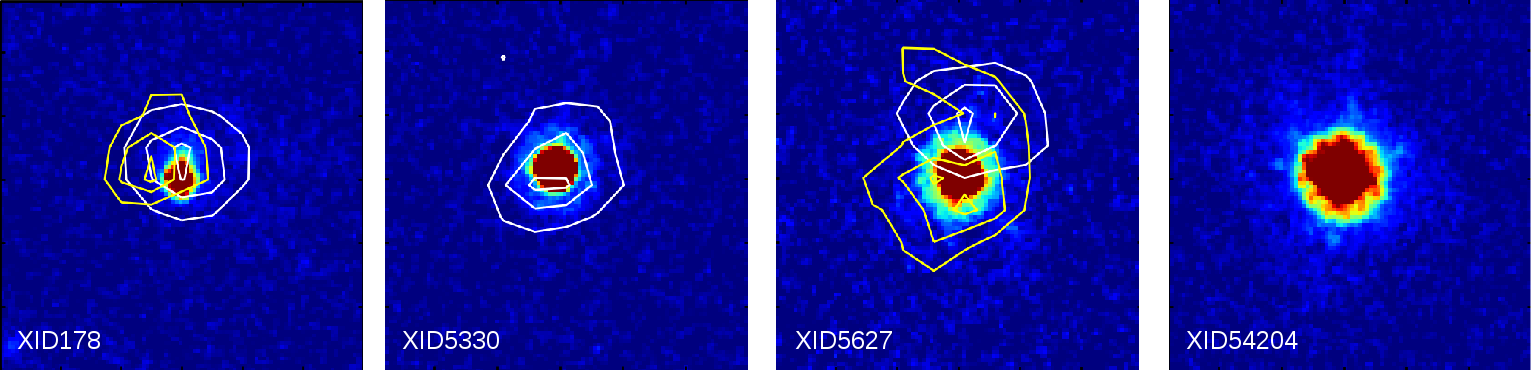}
\caption{3"$\times$3" HST-ACS images of the sample presented in
  this paper, with a spatial scale of 3"$\times$3" for each
  object. For XID178 and XID5627, the white contours represent the
  narrow component of the [\ion{O}{iii}] $\lambda$5007 emission line profile
  while the yellow contours represent the broad component while for
  XID5330, the white contours represent the single Gaussian component
  of [\ion{O}{iii}] $\lambda$5007.  All contours are at levels 50\%, 75\% and
  95\%. The SINFONI data are not corrected for
  astrometry. \label{hst_images}}
\end{figure*}

We looked at the Hubble Space Telescope (HST) images for these
objects, and the cutouts are shown in Fig. \ref{hst_images}. All
images have a spatial scale of
3$^{\prime\prime}\times$3$^{\prime\prime}$ which is comparable to the
flux and velocity maps presented in Sec. \ref{sec_results}. None of
these images show any disturbed morphology that could hint to a recent
merger, therefore we may disfavor the idea that the dynamical
signatures interpreted as outflows are actually due to mergers. 
  The white contours in Fig. \ref{hst_images} represent the narrow
  [\ion{O}{iii}] $\lambda$5007 emission for XID178, XID5330 and XID5627 while
  the yellow contours represent the broad component. It is clear from
  the figure that the ionized emission traces a larger fraction of the
  total gas in the host galaxy compared to the optical images.

Based on the star formation rates (SFR) of the host galaxies of our
sample of QSOs, we can also derive the kinetic power driven by
supernovae and stellar winds and compare it to the outflow power we
obtain to identify the possible source for such outflows. The power
driven due to these stellar processes can be calculated using
\cite{veilleux05} formalism:

\begin{equation}
\label{eq_power_sf}
P_{SF} = 7\cdot 10^{41}\times SFR  ~~\mathrm{M_{\odot}/yr}
\end{equation}

The power due to star formation for each of our targets are listed in
column 7 of Table \ref{outflow_properties} and can be compared with
the kinetic power of the outflow due to the ionized gas component
given in column 6 of the same table. For XID2028, \cite{cresci15}
reported a value of $\mathrm{\sim 1.5\times 10^{44} ~erg/s}$ for the
kinetic energy due to the ionized gas as traced by the H$\beta$ line,
while we obtain $\mathrm{\sim 5.8\times 10^{42} erg/s}$ from the
[\ion{O}{iii}] $\lambda$5007. The apparent discrepancy of a factor of 25 shows the difficulty of using [\ion{O}{iii}]  lines as 
tracer of the outflow mass as explained in Sec. \ref{sec_outflow_properties}. For reference, we use both the values we obtain using [\ion{O}{iii}] $\lambda$5007 and that obtained
by \cite{cresci15} using H$\beta$ in the plots to follow to
demonstrate the differences in power estimates from the two
lines. Note that all the mass outflow rates reported in Table
\ref{outflow_properties} are lower limits because additional gas phases are missing.

\begin{figure}
\centering
\includegraphics[scale=0.33]{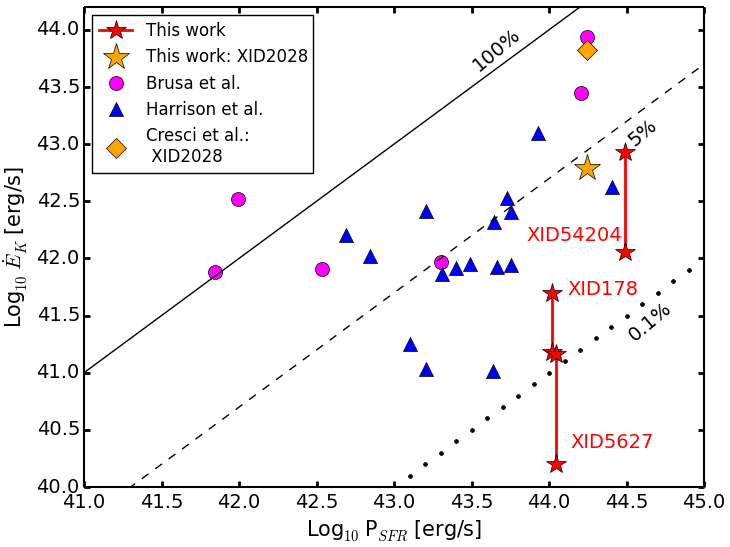}
\caption{Kinetic outflow power, $\dot{E}_{K}$ vs. the power driven by
  star formation, $P_{SFR}$. The solid, dashed and dotted lines
  represent 100\%, 5\% and 0.1\% ratios between the two powers. Our
  sample is represented by the star symbols while the other data
  points have been obtained from \cite{brusa15a, cresci15} and
  \cite{harrison14}. All data points represent outflow powers due to
  the ionized gas calculated from [\ion{O}{iii}] $\lambda$5007 line, except the
  data point from \cite[XID2028:][]{cresci15} which represents one of
  the objects from our selection and whose outflow properties were
  reported using H$\beta$ line. The two points for XID2028 show the
  differences in power estimates from the two emission lines. The
    range of kinetic power from our sample is due to the different
    prescriptions used to derive the outflow velocity.\label{psfr_pagn}}
\end{figure}

In Fig. \ref{psfr_pagn} we compare the outflow kinetic power with the
predicted energy input rate from star formation following Veilleux et
al. (2005) for our sample and other previous studies
\citep{harrison14, brusa15a, cresci15}. The outflow kinetic energy was
recomputed for each work from the literature to be consistent with our
assumption of electron density (100/cm$^{3}$) and temperature (10,000
K). The solid, dashed and dotted lines represent 100\%, 5\% and 0.1\%
ratios, respectively. For the two newly detected outflows (XID178 and
XID5627, red stars in the figure) having $> 0.1\%$ coupling between
the stellar energy input and the ISM is sufficient to make
star-formation viable as a dominant power source for the observed
outflows. For XID54204, star formation itself could not sustain
  the observed high velocities and kinetic power \citep{king11}.

\begin{figure}
\centering
\includegraphics[scale=0.35]{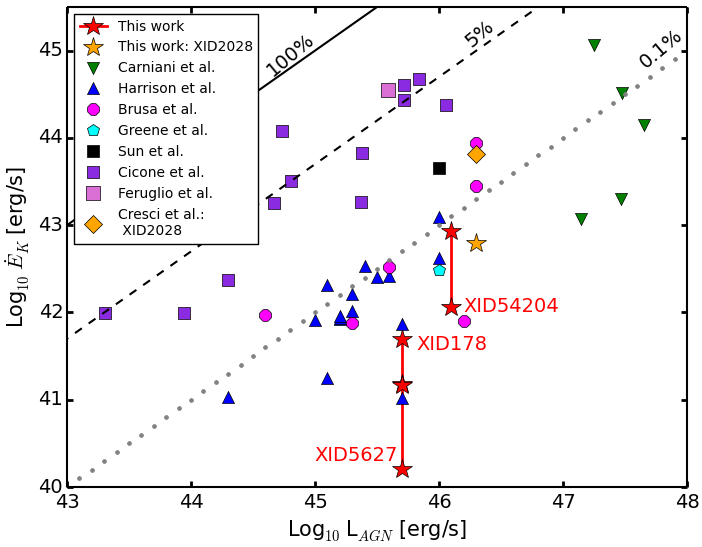}
\caption{Kinetic outflow power, $\mathrm{\dot{E}_{K}}$ (ionized and
  molecular) vs. bolometric luminosity of the AGN, L$_{AGN}$. The
  solid, dashed and dotted lines represent 100\%, 5\% and 0.1\% of the
  total bolometric luminosity of the AGN in the form of the outflow
  power. The data points from this work are shown in red stars
  (XID178, XID5627). The molecular outflows are represented by squares
  \citep{cicone14, feruglio15,sun14}, while the rest of the data
  points represent ionized outflows \citep{carniani15,
    harrison14,cresci15, brusa15a,greene12}. The outflow properties
  for all the ionized outflows were obtained using [\ion{O}{iii}] $\lambda$5007
  line except for \citealt{cresci15} for which analysis from H$\beta$
  is shown in this plot to represent the differences in analysis from
  the two lines for the same data.  \label{p_Lagn}}
\end{figure}

In Fig. \ref{p_Lagn} the outflow kinetic energy is plotted against the
AGN bolometric luminosity. We included in the figure other known
ionized (\citealt{greene12, harrison14, carniani15, brusa15a,
  cresci15}) and molecular (squares, \citealt{cicone14,
  sun14,feruglio15}) outflows. The dotted, dashed and the solid lines
represent the fraction of AGN bolometric luminosity in the form of the
outflow power (0.1\%, 5\% and 100\% respectively). Two of the
  objects presented in this work (XID178 and XID5627, red stars at
  lower bolometric luminosity) lie at the low energy end of the
  distribution of previously reported ionized outflows, while XID54204
  (red stars at a higher bolometric luminosity) falls in the regime of
  previously known ionized outflows. Theoretical models predict a
coupling of $\sim$0.1-5\% for AGN-driven outflows \citep{king05}, and
for these three objects a coupling less than $0.1\%$ of the radiative
power of the AGN with the ISM might be sufficient to power the
detected outflows. It is important to note that due to the multi-phase
nature of the outflows, our kinetic energy estimates are lower limit,
since one should consider the total mass of the outflow in the form of
ionized, molecular and neutral gas. This is supported by the fact that
the previously reported molecular outflows fall into a regime with
higher coupling ($\sim$5\%) between the outflow kinetic energy and
energy released by the AGN. The exact conversion factor to go from
ionized to the total gas mass is not known and might vary on an
object-by-object basis. Clearly follow-up observations at mm
wavelengths are needed to have a more complete picture.

Knowing the radius of the outflow for XID178 and XID5627 to be $\sim$8.6 kpc from the surface brightness profiles in Figs. \ref{178} and \ref{5627}, the dynamical time scale of the outflow
calculated from Eq. \ref{dynamic_time} comes out to be $\approx$ 18
Myr for both targets. This value is similar to previously reported
outflow timescales and AGN lifetimes (e.g. \citealt{cresci15,
  greene12,martini01}). This exercise could not be performed for
  XID54204 as we do not have the spatial resolution to determine the
  size of the outflow.
  
In summary both star-formation processes and AGN radiation could be
the dominant power source for the outflows presented in this
paper. Although, the observed high kinetic power for XID54204
  points to an AGN driven outflow, a higher S/N and better spatial
  resolution data is required to confirm this.

The outflow momentum rate, defined as $\mathrm{v_{o}\dot{M}_{o}}$, is
another fundamental parameter of the outflow which is crucial to
understand if it is momentum or energy conserving
\citep{zubovas12,faucher12}. Theoretical models predict that the
momentum of the kpc-scale outflow is boosted by the radiation pressure
of the wind from the AGN and the outflow energies are of the order of
$\approx\mathrm{20\cdot L_{AGN}/c}$ (AGN radiation pressure momentum)
implying an energy conserving outflow \citep{cicone14,
  feruglio15}. XID178 and XID5627 have similar bolometric luminosity,
outflow velocity and mass outflow rates (see Table
\ref{sample_selection} and \ref{outflow_properties}). The ratio of the
outflow momentum rate to that of AGN radiation pressure momentum comes
out to be much less than 1:1 ($\approx$ 0.01:1), which is also
consistent with previously reported ionized gas momentum rates
\citep{carniani15}. The lower ratio might be due to the discrepancy in
the outflow mass rates in the molecular and ionized gas phase.

\subsection{Uncertainties in the estimates}

Although there is substantial observational evidence that outflows are
common in AGNs, the derived quantitites associated to these outflows
are affected by significant uncertainties.

We have adopted the most commonly used outflow model, i.e. the
bi-conical outflow, to derive the mass outflow rates and hence the
power driven by the AGN. Mass outflow rate estimates would change if
we considered a shell-like outflow geometry instead. This has been
extensively discussed in \cite{cicone14} and \cite{maiolino12} where
it was shown that a shell-like geometry gives a higher outflow rate
(by a factor of $\sim$3) relatively to the multi-conical or spherical
geometry considered above.  Since we do not have a 3D view of these
galaxies, the exact outflow morphology of any of the targets could not
be inferred. However, with the 2D images in hand, one could assume a
bi-conical outflow model in all the objects with XID5627 probably
having a wide-angled outflow (inferred from the broad component map in
Fig. \ref{5627}, middle right panel) towards the line of sight of the
observer. This suggests that the outflow pattern might
vary on an object-by-object basis and might introduce errors due to
the assumptions made.

As we have illustrated in Figs. \ref{psfr_pagn} and \ref{p_Lagn},
  using different prescriptions for determining outflow velocities
  introduces a range of outflow properties, the extent of which
  depends on the emission line profile. In all cases,
  $\mathrm{w_{80}}$ gives a higher velocity value compared to the
  $\mathrm{v_{10}}$ on the integrated spectrum and the broad
  component. For a Gaussian profile $\mathrm{w_{80}}$ is close to
  the FWHM of the line and relates to the typical velocity of
  the emitting gas, while $\mathrm{v_{10}}$ gives an
  estimate of velocity the gas moving towards us at the high
  velocity end. The outflow model we use do not take into account
  these differences.

Another important source of uncertainty is the lack of a measurement
of the electron density for each single AGN studied. The range of
electron density in NLR is believed to be in the range $\le 50-1500$
$\mathrm{cm^{-3}}$ \citep{peterson97}. Observations of local galaxies
have shown that the electron density generally drops with distance
from the galactic centre, down to values $\la 50$ $\mathrm{cm^{-3}}$
in the galaxy outskirts \citep{bennert06}. The observed outflows in
our case are extended to kiloparsec scales, but
if the gas is dragged out by an over-pressurized bubble the density could be
relatively high. We adopt an average electron density of 100 $\mathrm{cm^{-3}}$, based on recent observations by \citealt{perna15} of a bright QSO (XID5321) using [\ion{S}{ii}] $\lambda$6716,31 for which an electron density of $\sim$120 $\mathrm{cm^{-3}}$ have been reported in the off-nuclear regions.
 However, this method is not effective in measuring
electron densities lower than 100 $\mathrm{cm^{-3}}$ (or greater than
1500 $\mathrm{cm^{-3}}$: \citealt{peterson97}). Moreover it is challenging to resolve the [\ion{S}{ii}] doublet near H$\alpha$ in many objects and is usually very faint. Hence, we have limitations
from observational point of view in determining the electron densities accurately.

A similar concern applies to the temperature estimates as well. The
normalization factors in Eqs. 3-5 could change by a factor of $\sim$10
depending on whether the temperature is assumed to be 10,000 K or
25,000 K since the emissivity changes by this factor.

Other sources of uncertainties come from using different lines for
deriving outflow properties (such as [\ion{O}{iii}]$\lambda$5007 and H$\beta$)
and the multi-phase nature of outflows which has been discussed
before. The overall errors in the estimates of mass outflow rates and
kinetic energies are therefore large, possibly of the order of a
factor $\sim 100$.

Hence, we stress the importance of further detailed observations to
constrain the outflow energies and hence accurately determine whether
AGN or the star formation could drive such outflows.

\section{Summary}    \label{sec_conclusions}

We summarize below the main results of this work:

\begin{itemize}
\item The selection of AGN in an outflowing phase based on the
  empirical curve by \cite{fabian08} (see Fig. \ref{forbidden_region})
  seems effective. We were able to verify the presence of such
    outflows using a kinematic analysis of the [\ion{O}{iii}] $\lambda$5007
    line in at least four out of a sample of five objects, three of
    them reported for the first time in this work (XID178, XID5627 and
    XID54204) and the fourth (XID2028) already presented in
    \cite{cresci15}.
\item In XID178 and XID5627, the outflow is extended up to
    $\sim$8.5 kpc in the host galaxy while for XID54204, the
    [\ion{O}{iii}] $\lambda$5007 emission is not spatially resolved. All the
    objects show high velocities of $\sim$500-1800 km/s in their
    integrated spectra. The spatial distribution of
    [\ion{O}{iii}] $\lambda$5007 in XID178 and XID5627 might suggest the
    presence of different outflow morphology in the galaxies.
\item For XID5330 we do not have enough S/N ratio and spectral
    resolution to trace an extended [\ion{O}{iii}] $\lambda$5007 emission (if
    present).
\item HST images of the sample do not show any disturbed morphology,
  possibly ruling out a merger-driven scenario for the observed
  outflows.
\item Based on the measured kinetic energies of the outflows, both
  star-formation ($\gtrsim 0.1\%$ coupling) and AGN radiation
  ($<0.1\%$ coupling) could be the dominant power source. In this work
  we were able to trace only the ionized phase of the
  outflows. Follow-up observations of the molecular components could
  allow us to put stronger constraints on the origin of these
  outflows.
\item The assumptions of current models introduce errors of several
  orders of magnitude on the outflow properties, which makes it
  difficult to infer the source of these outflows. Accurate
  determination of the electron density and electron temperatures in
  these galaxies is required to confirm if these outflows are powered
  by an AGN or star formation. Moreover, observations at mm
  wavelengths are required to trace the outflows component in
  molecular gas phase.
\end{itemize}

From this and previous works, it is now apparent that the presence of
outflows in high redshift galaxies hosting an AGN is very common. 
  But one has to be cautious since most of these studies are based on
  pre-selected targets, to maximize the chance to actually detect
  outflows. There is a strong need for a blind IFS survey of AGN to
trace the properties and the incidence of outflows as a function of
AGN physical properties. We have recently started a Large Program with
SINFONI at VLT, ``Survey for Unveiling the Physics and the Effect of
Radiative feedback'' (SUPER, PI Mainieri), designed to test the
presence of outflows and their impact on the host galaxy in a sample
of $>40$ AGNs at z$\sim$2 covering a wide range of bolometric
luminosities, Eddington ratios and star formation rates. Another
important goal of this survey will be to minimize the uncertainties
when computing the outflow energetics, in particular deep K-band
observations will allow reliable determinations of the electron
density object by object using the [\ion{S}{ii}]  doublets. Finally, such
survey should be complemented by similar observations in the mm regime
to characterize the molecular phase of the outflows.

%________________________________________________________________

\begin{acknowledgements}
  MB and GL acknowledge support from the FP7 Career Integration Grant
  "eEASy: supermassive black holes through cosmic time from current
  surveys to eROSITA-Euclid Synergies" (CIG 321913). The authors also
  acknowledge useful discussions with Michele Perna. We also thank the anonymous referee for her/his useful comments which helped us to improve the paper.
\end{acknowledgements}

%-------------------------------------------------------------------

\bibliographystyle{aa}
\bibliography{reference.bib}

\end{document}